\documentclass{aa}
\usepackage{epsfig,graphicx,natbib,url,twoopt}
\usepackage[varg]{txfonts}
\usepackage{hyperref}          %% for pdflatex
\usepackage[usenames]{color}
\usepackage{pdfcomment,acronym}      
\hypersetup{
  colorlinks=true,  
  urlcolor=blue,    
  linkcolor=red     
}

\def\RR #1\par{\noindent{\small \color{red} $\sharp$RR #1}\\[1ex]}
 
\bibpunct{(}{)}{;}{a}{}{,}    %% natbib cite format used by A&A and ApJ
\makeatletter
 \newcommandtwoopt{\citeads}[3][][]{%
   \nonstopmode%              %% fix to not stop at error message in latex
   \href{http://adsabs.harvard.edu/abs/#3}%
        {\def\hyper@linkstart##1##2{}%
         \let\hyper@linkend\@empty\citealp[#1][#2]{#3}}%   %% Rutten, 2000
   \biblink{#3}{\href{http://adsabs.harvard.edu/abs/#3}{ADS}}%
   \errorstopmode}            %% fix to resume stopping at error messages 
 \newcommandtwoopt{\citepads}[3][][]{%
   \nonstopmode%              %% fix to not stop at error message in latex
   \href{http://adsabs.harvard.edu/abs/#3}%
        {\def\hyper@linkstart##1##2{}%
         \let\hyper@linkend\@empty\citep[#1][#2]{#3}}%     %% (Rutten 2000)
   \biblink{#3}{\href{http://adsabs.harvard.edu/abs/#3}{ADS}}%
   \errorstopmode}            %% fix to resume stopping at error messages
 \newcommandtwoopt{\citetads}[3][][]{%
   \nonstopmode%              %% fix to not stop at error message in latex
   \href{http://adsabs.harvard.edu/abs/#3}%
        {\def\hyper@linkstart##1##2{}%
         \let\hyper@linkend\@empty\citet[#1][#2]{#3}}%     %% Rutten (2000)
   \biblink{#3}{\href{http://adsabs.harvard.edu/abs/#3}{ADS}}%
   \errorstopmode}            %% fix to resume stopping at error messages 
 \newcommandtwoopt{\citeyearads}[3][][]{%
   \nonstopmode%              %% fix to not stop at error message in latex
   \href{http://adsabs.harvard.edu/abs/#3}%
        {\def\hyper@linkstart##1##2{}%
         \let\hyper@linkend\@empty\citeyear[#1][#2]{#3}}%  %% 2000
   \biblink{#3}{\href{http://adsabs.harvard.edu/abs/#3}{ADS}}%
   \errorstopmode}            %% fix to resume stopping at error messages 
\makeatother

\newcount\longrefs
\def\aap{\ifnum\longrefs=1 {Astron.\ Astrophys.}\else 
                           {A\hbox{\rm \&}A}\fi}
\def\aapr{\ifnum\longrefs=1 {Astron.\ Astrophys.\ Rev.}\else 
                            {A\hbox{\rm \&}AR}\fi}
\def\aaps{\ifnum\longrefs=1 {Astron.\ Astrophys.\ Suppl.}\else 
                            {A\hbox{\rm \&}A Suppl.}\fi}
\def\actaa{\ifnum\longrefs=1 {Acta Astronomica}\else
                            {Acta Astron.}\fi}
\def\aipcs{\ifnum\longrefs=1 {Am.\ Inst.\ Phys.\ Conf.\ Series}\else
                             {AIP Conf.\ Ser.}\fi}
\def\aj{\ifnum\longrefs=1 {Astron.\ J.}\else 
                          {AJ}\fi} 
\def\ao{\ifnum\longrefs=1 {Applied Optics}\else 
                           {Appl.\ Opt.}\fi} 
\def\aspcs{\ifnum\longrefs=1 {Astron.\ Soc.\ Pacific Conf.\ Series}\else 
                           {ASP Conf.\ Ser.}\fi} 
\def\apj{\ifnum\longrefs=1 {Astrophys.\ J.}\else 
                           {ApJ}\fi} 
\def\apjl{\ifnum\longrefs=1 {Astrophys.\ J. Lett.}\else 
                            {ApJL}\fi} 
\def\aplett{\ifnum\longrefs=1 {Astrophys.\ J. Lett.}\else 
                            {ApJ}\fi} 
\def\apjs{\ifnum\longrefs=1 {Astrophys.\ J. Suppl.}\else 
                            {ApJS}\fi}
\def\apss{\ifnum\longrefs=1 {Astrophys.\ and Space Science}\else 
                            {Astrophys.\ Space Sci.}\fi}
\def\araa{\ifnum\longrefs=1 {Ann.\ Rev.\ Astron.\ Astrophys.}\else 
                            {ARA\hbox{\rm \&}A}\fi}
\def\azh{\ifnum\longrefs=1 {Astronomicheskii Zhurnal}\else 
                            {Astron.\ Zhur.}\fi}
\def\baas{\ifnum\longrefs=1 {Bull.\ Am.\ Astron.\ Soc.}\else 
                            {BAAS}\fi}
\def\bain{\ifnum\longrefs=1 {Bull.\ Astronom.\ Institutes Netherlands}\else
                            {Bull.\ Astr.\ Inst.\ Neth.}\fi}
\def\cjaa{\ifnum\longrefs=1 {Chinese Jour.\ Astron.\ Astrophys.}\else 
                            {Chin.\ J.\ A\&A}\fi}
\def\gca{\ifnum\longrefs=1 {Geochim.\ Cosmochim.\ Acta}\else 
                           {Geochim.\ Cosmochim.\ Acta}\fi}
\def\grl{\ifnum\longrefs=1 {Geophys.\ Res.\ Lett.}\else 
                           {Geoph.\ Res.\ Lett.}\fi}
\def\iaucirc{\ifnum\longrefs=1 {IAU Circulars}\else 
                          {IAU Circ.}\fi}
\def\icarus{\ifnum\longrefs=1 {Icarus}\else 
                          {Icarus}\fi}
\def\ip{\ifnum\longrefs=1 {in press}\else 
                          {in press}\fi}
\def\jcap{\ifnum\longrefs=1 {Jour.\ Cosmology Astropart.\ Phys.}\else 
                          {JCAP}\fi}
\def\jgr{\ifnum\longrefs=1 {J.\ Geophys.\ Res.}\else 
                           {J.\ Geophys.\ Res.}\fi}  
\def\jrasc{\ifnum\longrefs=1 {J.\ Royal Astron.\ Soc.\ Canada}\else 
                           {JRAS Can.}\fi}  
\def\memsai{\ifnum\longrefs=1 {Mem.~Soc.~Astron.~Italiana}\else
                              {MmSAI}\fi}
\def\mnras{\ifnum\longrefs=1 {Mon.\ Not.\ Roy.\ Astron.\ Soc.}\else 
                             {MNRAS}\fi} 
\def\na{\ifnum\longrefs=1 {New Astronomy}\else 
                           {New Astron.}\fi}
\def\nar{\ifnum\longrefs=1 {New Astronomy rev.}\else 
                           {New Astron.\ Rev.}\fi}
\def\nat{\ifnum\longrefs=1 {Nature}\else 
                           {Nat}\fi}
\def\pasa{\ifnum\longrefs=1 {Pub.\ Astron.\ Soc.\ Australia}\else 
                            {PASA}\fi} 
\def\pasj{\ifnum\longrefs=1 {Pub.\ Astron.\ Soc.\ Japan}\else 
                            {PASJ}\fi} 
\def\pasp{\ifnum\longrefs=1 {Pub.\ Astron.\ Soc.\ Pacific}\else 
                            {PASP}\fi} 
\def\physscr{\ifnum\longrefs=1 {Physica Scripta}\else 
                            {Phys.\ Scrip.}\fi} 
\def\planss{\ifnum\longrefs=1 {Planetary \& Space Science}\else 
                            {Plan. \& Space Sci.}\fi} 
\def\procspie{\ifnum\longrefs=1 {Proc.\ SPIE}\else 
                            {Proc.\ SPIE}\fi} 
\def\qjras{\ifnum\longrefs=1 {Quarterly J.\ Royal Astron.\ Soc.}\else 
                            {QJRAS}\fi} 
\def\rmxaa{\ifnum\longrefs=1 {Revista Mexicana de Astron.\ y Astrofys.}\else 
                            {RMxAA}\fi} 
\def\sa{\ifnum\longrefs=1 {Soviet Astron..}\else 
                               {Sov.\ Astron.}\fi}
\def\skytel{\ifnum\longrefs=1 {Sky \& Telescope}\else 
                            {Sky \& Tel.}\fi} 
\def\solphys{\ifnum\longrefs=1 {Solar Phys.}\else 
                               {SoPh}\fi}
\def\sovast{\ifnum\longrefs=1 {Soviet Astronomy}\else 
                               {Sov.\ Ast.}\fi}
\def\ssr{\ifnum\longrefs=1 {Space Science Rev.}\else 
                               {Space\ Sci.\ Rev.}\fi}
\def\zap{\ifnum\longrefs=1 {Zeitschr.\ f.\ Astrophysik}\else
                               {Z.\ Astrophys.}\fi}

\makeatletter
\newcommand{\bibnote}[2]{\@namedef{#1note}{#2}}
\newcommand{\biblink}[2]{\@namedef{#1link}{#2}}
\makeatother

\def\wlRRtop#1{\href{/http://www.staff.science.uu.nl/~rutte101}{#1}}
\def\wlsolarsoft#1{\href{http://www.lmsal.com/solarsoft}{#1}}

\def\acdef#1{\acl{#1} ({#1})}     %RR to avoid \ac first-use confusion
\newacro{AA}{Astronomy \& Astrophysics}  %RR I can't define A&A
\newacro{ADS}{Astrophysics Data System}
\newacro{AIA}{Atmospheric Imaging Assembly}
\newacro{ALMA}{Atacama Large Millimeter/submillimeter Array}
\newacro{AO}{adaptive optics}
\newacro{ApJ}{Astrophysical Journal}
\newacro{AR}{active region}
\newacro{BFI}{Broad-band Filter Imager}
\newacro{CE}{coronal equilibrium}
\newacro{CfA}{Center for Astrophysics}
\newacro{CME}{coronal mass ejection}
\newacro{CRD}{complete redistribution}
\newacro{CRISP}{CRisp Imaging SpectroPolarimeter}
\newacro{CRISPEX}{CRisp SPectral EXplorer}
\newacro{CS}{coherent scattering}
\newacro{DEM}{Differential Emission Measure}
\newacro{DF}{dynamic fibril}
\newacro{DKIST}{Daniel K. Inouye Solar Telescope}
\newacro{DLR}{Deutsches Zentrum f\"ur Luft- und Raumfahrt}
\newacro{DOT}{Dutch Open Telescope}
\newacro{DST}{Richard B. Dunn Solar Telescope}   %RR or Domeless ST at Hida
\newacro{EB}{Ellerman bomb}
\newacro{EDP}{\'{E}dition Diffusion Presse}  %RR they say so
\newacro{EIT}{Extreme ultraviolet Imaging Telescope}
\newacro{EPIC}{European participation in Solar-C}
\newacro{ERC}{European Research Council}
\newacro{ESA}{European Space Agency}
\newacro{EST}{European Solar Telescope}
\newacro{EUV}{extreme ultraviolet}
\newacro{FAF}{flaring active-region fibril}
\newacro{FITS}{Flexible Image Transport System}
\newacro{FOV}{field of view}
\newacro{fov}{field of view}
\newacro{FWHM}{full width at half maximum}
\newacro{HAO}{High Altitude Observatory}
\newacro{HD}{hydrodynamics}
\newacro{Hi-C}{High Resolution Coronal Imager Sounding Rocket}
\newacro{HMI}{Helioseismic and Magnetic Imager}
\newacro{IAA}{Instituto de Astrof\'{i}sica de Andaluc\'{i}a}
\newacro{IAC}{Instituto de Astrof\'{i}sica de Canarias}
\newacro{IAS}{Institut d'Astrophysique Spatiale}
\newacro{IBIS}{Interferometric Bi-dimensional Spectrometer}
\newacro{IDL}{Interactive Data Language}
\newacro{IMaX}{Imaging Magnetograph eXperiment}
\newacro{INAF}{Istituto Nazionale di Astrofisica}
\newacro{IB}{IRIS bomb}
\newacro{IR}{infrared}
\newacro{IRIS}{Interface Region Imaging Spectrograph}
\newacro{ISAS}{Institute of Space and Astronautical Science}
\newacro{ISP}{Institute for Solar Physics}
\newacro{ISS}{International Space Station}
\newacro{ISSI}{International Space Science Institute}
\newacro{ITA}{Institute for Theoretical Astrophysics}
\newacro{JAXA}{Japan Aerospace Exploration Agency}
\newacro{KIS}{Kiepenheuer--Institut f\"{u}r Sonnenphysik}
\newacro{KPNO}{Kitt Peak National Observatory}
\newacro{LASP}{Laboratory for Atmospheric and Space Physics}
\newacro{LC}{liquid cristal}
\newacro{LMSAL}{Lockheed Martin Solar and Astrophysics Labratory}
\newacro{LOS}{line of sight}
\newacro{LTE}{local thermodynamic equilibrium}
\newacro{MC}{magnetic concentration}
\newacro{MCAO}{multi-conjugate adaptive optics} 
\newacro{MDI}{Michelson Doppler Imager}
\newacro{ME}{Milne-Eddington} 
\newacro{MHD}{magnetohydrodynamics}
\newacro{MOMFBD}{Multi-Object Multi-Frame Blind Deconvolution}
\newacro{MPE}{Max--Planck--Institut f\"ur extraterrestrische Physik}
\newacro{MPG}{Max--Planck--Gesellschaft}
\newacro{MPS}{Max Planck Institute for Solar System Research}
\newacro{MSSL}{Mullard Space Science Laboratory}
\newacro{MTF}{modulation transfer function}
\newacro{NAOJ}{National Astronomical Observatory of Japan}
\newacro{NASA}{National Aeronautics and Space Administration}
\newacro{NLTE}{non-local thermodynamic equilibrium}
\newacro{NLFFF}{non-linear force-free field}
\newacro{NOAA}{National Oceanic and Atmospheric Administration}
\newacro{non-E}{non-equilibrium}
\newacro{NSO}{National Solar Observatory}
\newacro{NWO}{Netherlands Organisation for Scientific Research}
\newacro{PHE}{propagating heating event}
\newacro{PRD}{partial redistribution}
\newacro{PROBA2}{PRoject for OnBoard Autonomy}
\newacro{PSBE}{post Saha-Boltzmann extinction}
\newacro{PSF}{point spread function}
\newacro{QS}{quiet Sun}
\newacro{QSEB}{quiet-Sun Ellerman-like brightening} 
\newacro{RAL}{Rutherford Appleton Laboratory}
\newacro{RBE}{rapid blue-shifted excursion}
\newacro{R-MHD}{radiation hydrodynamics}
\newacro{rms}{root mean square}
\newacro{RMS}{root mean square}
\newacro{ROB}{Royal Observatory of Belgium}
\newacro{ROI}{region of interest}
\newacro{RRE}{rapid red-shifted excursion}
\newacro{RTE}{radiative transfer equation}
\newacro{SE}{statistical equilibrium}
\newacro{SB}{Saha Boltzmann}
\newacro{SDO}{Solar Dynamics Observatory}
\newacro{SJI}{slit-jaw image}
\newacro{SNR}{signal-to-noise ratio}
\newacro{SO}{Solar Orbiter}
\newacro{SoHO}{Solar and Heliospheric Observatory}
\newacro{SP}{Spectropolarimeter}
\newacro{SST}{Swedish 1-m Solar Telescope}
\newacro{SUMER}{Solar Ultraviolet Measurements of Emitted Radiation}
\newacro{SUFI}{Sunrise Filter Imager}
\newacro{SVD}{singular value decomposition}
\newacro{SVST}{Swedish Vacuum Solar Telescope}
\newacro{THEMIS}{T\'{e}lescope H\'{e}liographique pour l'Etude du 
   Magn\'{e}tisme et des Instabilit\'{e} Solaires}     %RR Wow
\newacro{TR}{transition region}
\newacro{TRACE}{Transition Region and Coronal Explorer}
\newacro{TSI}{total solar irradiance}
\newacro{UT}{Universal Time}
\newacro{UV}{ultraviolet}
\newacro{VAULT}{Very high angular resolution ultraviolet telescope}
\newacro{VIRGO}{Variability of solar IRradiance and Gravity Oscillations}
\newacro{VTT}{Vacuum Tower Telescope}    %RR Germans say "vauteetee"
\newacro{XRT}{X-Ray Telescope}

\def\acp#1{\pdftooltip{\acs{#1}}{\acl{#1}}} %% pdfcomment.sty Apr 26 2015 

\def\nl{,\ } %%\def\nl{\newline}  %% redefine as \newline for mail addresses

\def\ITA{Institute of Theoretical Astrophysics\nl
         University of Oslo\nl
         P.O. Box 1029, Blindern\nl N-0315 Oslo\nl Norway}

\def\LA{Lingezicht Astrophysics\nl 't Oosteneind 9\nl 4158\,CA Deil\nl 
        The Netherlands}

\long\def\startignore #1\stopignore{}   %% use \startignore....\stopignore

\def\rmit#1{{\it #1}}              %% italics (RR style, Kluwer)
\def\etal{\rmit{et al.}}           %% use \etal\ for space behind it        
           
\def\eg{\rmit{e.g.,}}              %% , required American (Webster 1681)
\def\cf{cf.}                       %% no Latin, always Roman (Webster 1686)

\def\specchar#1{\uppercase{#1}}    %% redefine for A&A, small caps
\def\FeI{\mbox{Fe\,\specchar{i}}} 
\def\FeII{\mbox{Fe\,\specchar{ii}}} 
 
\def\FeIX{\mbox{Fe\,\specchar{ix}}}

\def\FeXII{\mbox{Fe\,\specchar{xii}}}

\def\HI{\mbox{H\,\specchar{i}}} 
 
\def\HeII{\mbox{He\,\specchar{ii}}}

\def\SiIV{\mbox{Si\,\specchar{iv}}}

\def\Halpha{\mbox{H\hspace{0.1ex}$\alpha$}} %% \Halpha\ for space behind it

\def\Lyalpha{\mbox{Ly$\hspace{0.2ex}\alpha$}}

\def\CaIR{\mbox{Ca\,\specchar{ii}\,8542\,\AA}} 

\def\level #1 #2#3#4{$#1 \; ^{#2} \mbox{#3} ^{#4}$}   

\def\rme{{\rm e}}

 \def\rmH{{\rm H}}

\def\deg{\hbox{$^\circ$}}       %% \def for overwriting, \box for math

\def\kms{\hbox{km$\;$s$^{-1}$}}

\def\tis{\!=\!}                            %% tighter spacing
\def\tapprox{\!\approx\!}                  %% tighter spacing
\def\rmit#1{#1}                 %% A&A & ApJ: latin abbreviations in Roman
\def\specchar#1{{\textsc{#1}}}  %% small caps for A&A

\def\revpar{{{\!\boldmath$\forall$\,}}} %% revision delete marker
\def\revpar{}                 %% ignore revision delete marker
\long\def\rev#1{{{\bf #1}}}   %% revision change/addition boldface
\long\def\rev#1{#1}           %% ignore revision change/addition boldface

\def\PubI{\href{http://adsabs.harvard.edu/abs/2016A&A...590A.124R}{Pub\,1}}
\bibnote{2016A&A...590A.124R}{(Pub\,1)}  

\begin{document}

\title{H-alpha features with hot onsets. II. A contrail fibril}
\subtitle{}

\author{R. J. Rutten\inst{1,2}
  \and
  L. H. M. Rouppe van der Voort\inst{2}
}

\institute{\LA \and \ITA} 

\date{Received 15 October 2015 /  Accepted 22 September 2016}

\abstract{The solar chromosphere observed in \Halpha\ consists mostly
of narrow fibrils. 
The longest typically originate in network or plage and arch far over
adjacent internetwork. 
We use data from multiple telescopes to analyze one well-observed
example in a quiet area.
It resulted from the earlier passage of an accelerating disturbance in
which the gas was heated to \revpar \rev{high} temperature as in the
spicule-II phenomenon. 
After this passage a dark \rev{\Halpha} fibril appeared \revpar as
a contrail. \revpar
We use Saha-Boltzmann extinction estimation to gauge the onset
\rev{and subsequent} visibilities in various diagnostics and conclude
that such \revpar \Halpha\ fibrils can indeed be contrail phenomena,
not indicative of the thermodynamic and magnetic environment when they
are observed but of more dynamic \revpar happenings \rev{before}.
\rev{They} do not connect \revpar across internetwork cells but
represent \rev{launch} tracks of \rev{heating events} \revpar \rev{and}
chart magnetic field \revpar \rev{during} launch, not at present. }

\keywords{Sun: activity -- Sun: atmosphere -- Sun: magnetic fields}

\maketitle

%%%%%%%%%%%%%%%%%%%%%%%%%%%%%%%%%%%%%%%%%%%%%%%%%%%%%%%%%%%%%%%%%%%%%%%%%%%%
\section{Introduction}\label{sec:introduction}
%%%%%%%%%%%%%%%%%%%%%%%%%%%%%%%%%%%%%%%%%%%%%%%%%%%%%%%%%%%%%%%%%%%%%%%%%%%%
\revpar \rev{In the Balmer \Halpha\ line at 6563\,\AA\ } \revpar the solar
chromosphere shows canopies of long thin features called fibrils.
They occur everywhere where there is some magnetic activity. \revpar
Understanding the chromosphere necessitates understanding these.
\revpar \rev{Here we show a case where this requires} identifying
preceding events \rev{and suggest that} not doing so resembles
studying jet contrails in our sky without appreciating they were made
by aircraft.

The literature on \Halpha\ fine structure is vast (\eg\ the landmark
thesis of \citeads{1964PhDT........83B} 
and the review by
\citeads{1974soch.book.....B}). 
We therefore limit \rev{our} introductory summary to various types of
\Halpha\ fibrils and only to recent and pertinent results.

\paragraph{Dynamic fibrils.}
%%%%%%%%%%%%%%%%%%%%%%%%
The name was given by
\citetads{2006ApJ...647L..73H} 
and \citetads{2007ApJ...655..624D} 
who after decades of slack progress since Beckers' thesis
identified and explained a particular \Halpha\ fibril type.
\rev{These} \revpar are fairly short dark fibrils that jut out in phased rows
from plage and active network, \revpar
\rev{representing} field-guided acoustic shock waves sloshed up by
the $p$-mode interference pattern into relatively tenuous magnetic
field bundles. \revpar
\rev{Short dynamic fibrils} \revpar are \revpar \rev{similar} \revpar
but \rev{occur} in sunspot chromospheres \revpar
(\citeads{2013ApJ...776...56R}; 
\revpar
\citeads{2014ApJ...787...58Y}). 

\paragraph{Spicules-II, RBEs, RREs.}
%%%%%%%%%%%%%%%%%%%%%%%%%%%%%%%%
Spicules-II \revpar discovered by
\citetads{2007PASJ...59S.655D} 
\rev{are} long, slender, highly dynamic jets jutting out from the
limb, \revpar \rev{with much} dynamics \revpar attributed to \revpar
\rev{multiple} Alfv\'enic wave modes (\eg\ De~Pontieu \etal\
\citeyearads{2007Sci...318.1574D}, 
\citeyearads{2012ApJ...752L..12D}; 
\citeads{2008ApJ...673L.219M}). 
\revpar Their manifestations on the disk are slender jet-like
Doppler-shifted features observed in the wings of \Halpha\ and \CaIR\
\revpar \rev{as} rapid blueshifted excursions (RBE) and rapid
redshifted excursions (RRE)
(\citeads{2008ApJ...679L.167L}; 
\citeads{2009ApJ...705..272R}; 
Sekse \etal\
\citeyearads{2012ApJ...752..108S}, 
\citeyearads{2013ApJ...769...44S}). 
\revpar Their tops \revpar get very hot
(\citeads{2011Sci...331...55D}; 
\rev{\citeads{2014ApJ...792L..15P};} 
\citeads{2015ApJ...799L...3R}; 
\citeads{2015ApJ...806..170S}). 
\rev{They} \revpar tend to arise from plage and active network in
areas that are not too active and mainly unipolar (\eg\ Fig.~2 of
\citeads{2009ApJ...705..272R}) 
\rev{and} \revpar favor quiet Sun and coronal holes
(\citeads{2012ApJ...759...18P}). 

\paragraph{Long fibrils.}
%%%%%%%%%%%%%%%%%%%%%%
With this name we \rev{denote} the \revpar \rev{ubiquitous} slender
\Halpha\ structures that jut out from network and plage and reach far
out over adjacent internetwork, often giving the impression of
spanning from one side of a supergranulation cell to another \revpar
\rev{tracing} closed-field \revpar canopies.
\revpar We suggest below that \rev{this interpretation is not correct}
\revpar and therefore \revpar do not want to call them ``long
closed-loop fibrils'', nor ``long internetwork fibrils'' because they
invariably are rooted in network or plage at least on one
side. \revpar We thought about ``long arching fibrils'' because they
are generally curved, more so than dynamic fibrils, but eventually
chose to simply \rev{call them} long fibrils. 

They are \revpar absent only in the very quietest areas
(\citeads{2007ApJ...660L.169R}), 
but \revpar prominently cover the solar surface above active regions
and widely around them. 
\rev{Overall} they constitute the chromosphere
as \rev{named} by \citetads{1868RSPS...17..131L}. 
\revpar Their nature \rev{remains unclear}, nor whether they actually
map magnetic fields as one naturally supposes when viewing their
patterns. 
\revpar
They \revpar are often \rev{interpreted} as cylindrical flux tubes
(\eg\ \citeads{1971SoPh...20..298F}; 
\citeads{2008A&A...486..577S}), 
but also \revpar as sheets or warps in sheets
(\citeads{2011ApJ...730L...4J}; 
%% \citeads{2012ApJ...751...75G}; 
\citeads{2014ApJ...785..109L}) 
or as density corrugations
(Leenaarts \etal\
\citeyearads{2012ApJ...749..136L}, 
\citeyearads{2015ApJ...802..136L}). 

%% fig:full
%===========================================================================
\begin{figure*}
  \centerline{\includegraphics[width=\textwidth]{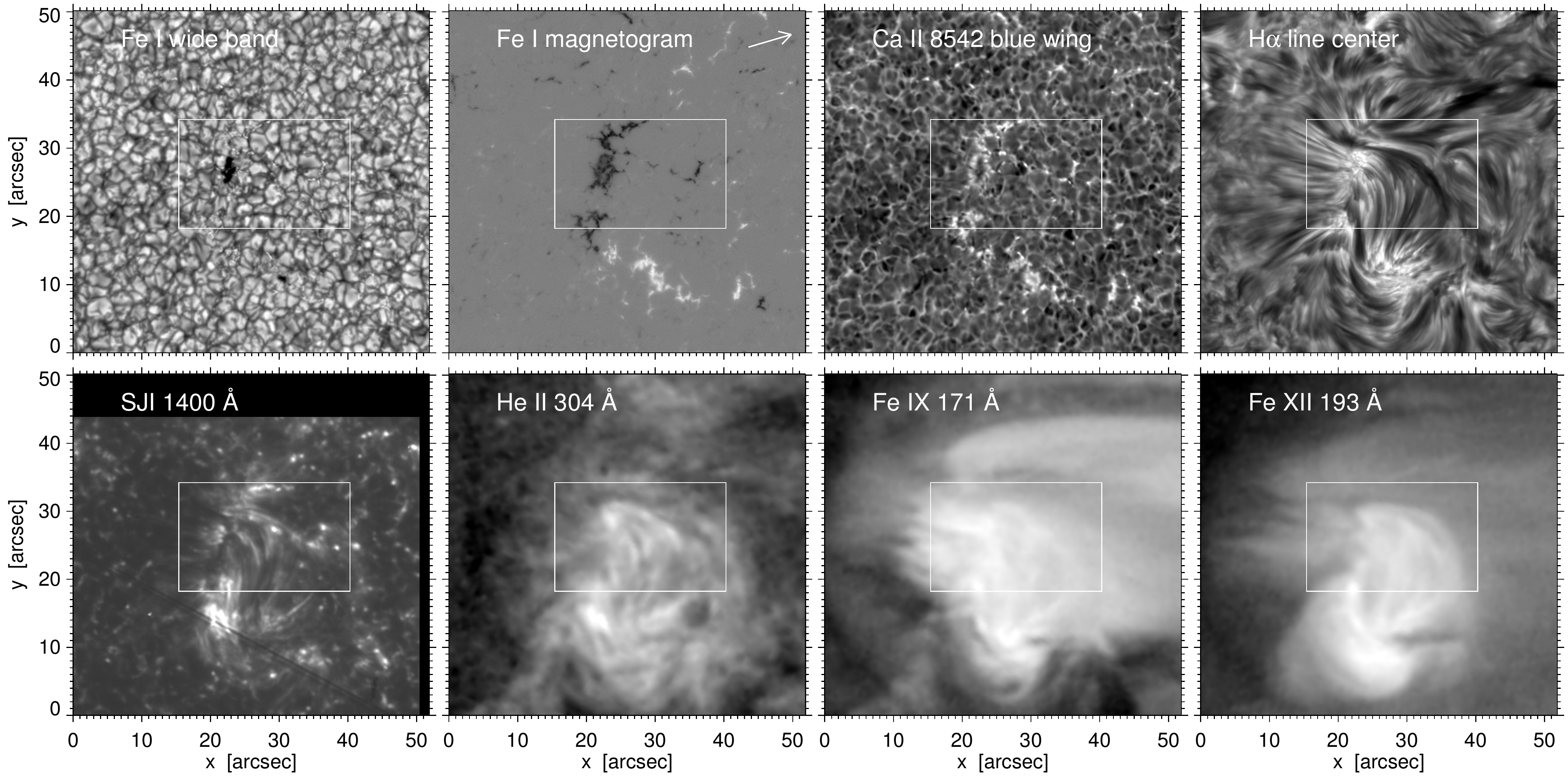}}
    \caption[]{\label{fig:full} 
    Overview of the observed area at 08:22:23\,UT.
    The field of view was centered at solar
    $(X,Y)\tis(-202, 206)$~arcsec with viewing angle $\mu\tis0.95$.
    It is clockwise rotated with respect to $(X,Y)$ over 62.6\deg.
    The arrow in the second panel points to disk center.
    {\em Upper row\/}: \acp{SST} images: \FeI\,6303\,\AA\ wide band,
    \FeI\,6303\,\AA\ magnetogram, blue wing of \CaIR\ at
    $\Delta \lambda \tis -0.7$\,\AA\ from line center, \Halpha\ line
    center.
    {\em Lower row\/}: corresponding images from \acp{IRIS} and
    \rev{\acp{SDO}/}\acp{AIA}: 1400\,\AA\ slitjaw, \HeII\,304,\,\AA\,
    \FeIX\,171\,\AA, \FeXII\,193\,\AA.  
    The pixel sizes are 0.057\,arcsec for the \acp{SST}, 0.167\,arcsec
    for \acp{IRIS}, 0.6\,arcsec for \acp{AIA}.
    The white frame defines the subfield used in
    Fig.~\ref{fig:sequence}.
    In the fourth panel it contains the long black arching fibril
    discussed \rev{here}.
    }
\end{figure*}
%===========================================================================

%% fig:sequence
%===========================================================================
\begin{figure*}
  \centerline{\includegraphics[width=\textwidth]{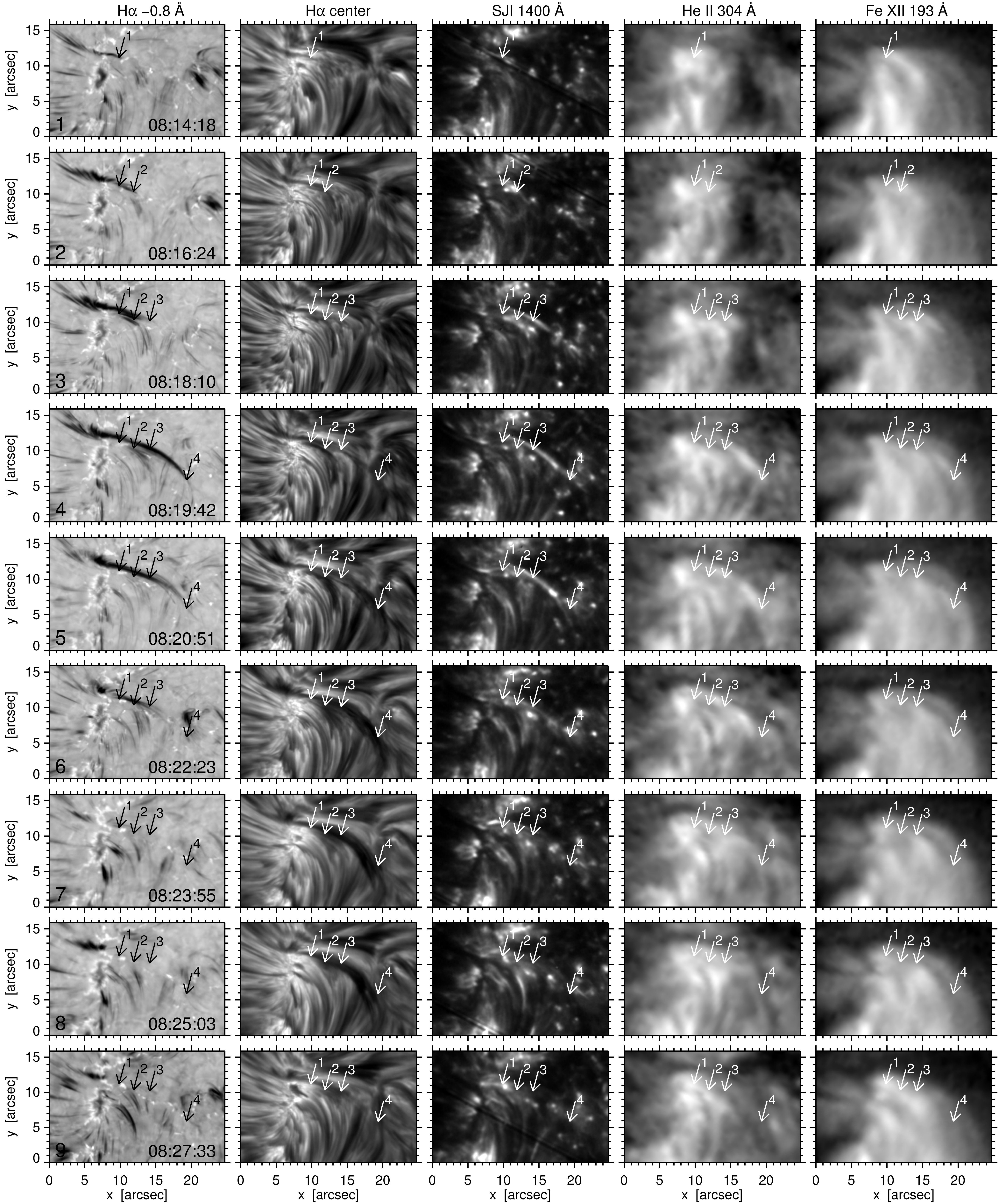}}
    \caption[]{\label{fig:sequence} 
    Time sequence of image cutouts defined by the subfield frame in
    Fig.~\ref{fig:full}. 
    \rev{The row number and the observing moment of each row are}
    specified in the first panel \rev{of each row}.
    {\em Columns\/}: \Halpha\ images at
    $\Delta \lambda \tis -0.8$~\AA\ from line center (\acp{SST}),
    \Halpha\ line center (\acp{SST}), 1400\,\AA\ slitjaw (\acp{IRIS}),
    \HeII\,304\,\AA\ (\acp{AIA}), \FeXII\,193\,\AA\ (\acp{AIA}).
    \rev{The four numbered arrows mark the location of the tip of the dark
    extending precursor streak defined successively for the
    first four \Halpha\ blue-wing panels.}  
    Each frame is bytescaled individually. 
    Row 6 corresponds to Fig.~\ref{fig:full}.
    The \acp{IRIS} slit is visible in the top two and bottom two
    1400\,\AA\ panels.
    }
\end{figure*}
%===========================================================================

%% fig:haca
%===========================================================================
\begin{figure*}
  \centerline{\includegraphics[width=\textwidth]{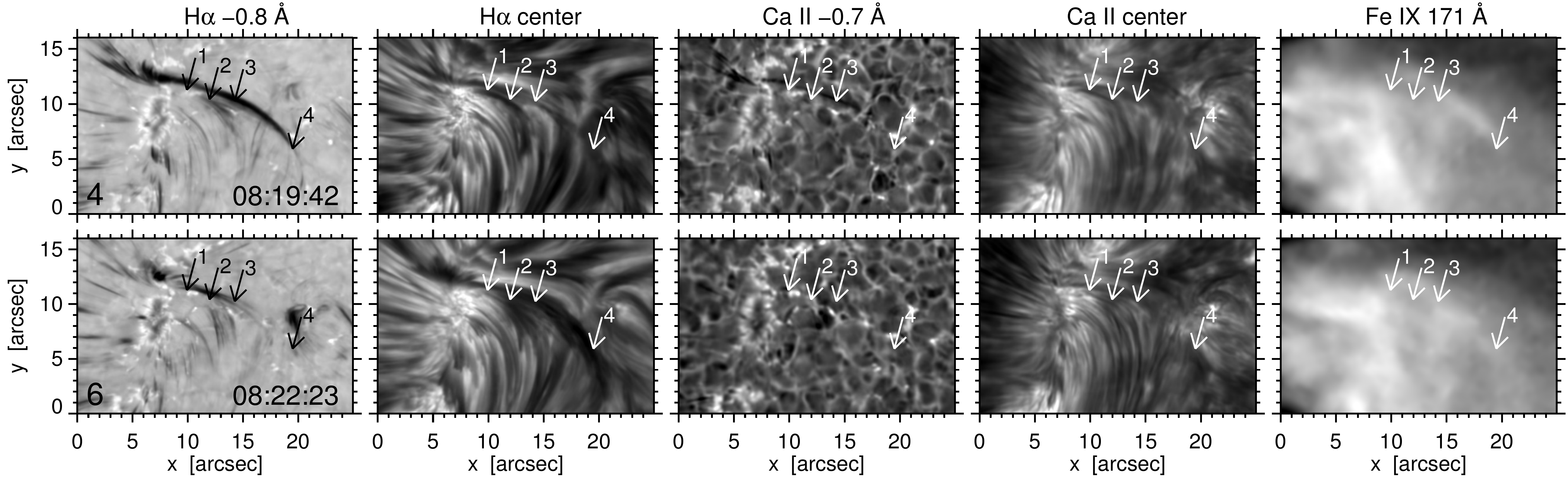}}
  \caption[]{\label{fig:haca} 
  Image cutouts as in Fig.~\ref{fig:sequence}.
  The first two columns repeat the first panels of rows 4 and 6 in
  Fig.~\ref{fig:sequence}.
  The third and fourth columns contain corresponding cutouts of \CaIR,
  in the blue wing at $\Delta \lambda \tis -0.7$\,\AA\ from line
  center and at line center. 
  The last column contains corresponding cutouts from \FeIX\,171\,\AA\
  (\acp{AIA}).
  }
\end{figure*}
%===========================================================================

\revpar Here we detail the formation of a single long \Halpha\ fibril
which we deem potentially exemplary for the \revpar class.
\rev{Our} study was triggered by \rev{a study} \revpar of Ellerman
bomb \rev{visibilities} 
%%!! (\citealp{2016A&A...590A.124R}, 
(\citeads{2016A&A...590A.124R}, 
henceforth Pub\,1) which \revpar was inspired by the 2D non-equilibrium
\acp{MHD} simulation of
\citetads{2007A&A...473..625L} 
\revpar called HION henceforth following \PubI.
The effects of dynamic hydrogen ionization in the HION atmosphere
defined the following recipe to understand marked presence of \Halpha\
in and after dynamical instances with hot and dense onsets: (1)
evaluate the \Halpha\ extinction coefficient \revpar (\HI\ $n\tis2$
level population) during the onset by assuming the Saha-Boltzmann
(\revpar SB) value, (2) use the resulting large population also for
cooler surrounding gas in reach of scattering \Lyalpha\ radiation from
the hot structure, and (3) maintain this large population subsequently
during \rev{cooling and recombination}.

\rev{This post-Saha-Boltzmann-extinction (PSBE)} recipe \revpar holds
for shocks in the HION atmosphere and was applied successfully to
Ellerman bombs in \PubI, \rev{including the suggestion that} \revpar
it \rev{applies} partially even to the \SiIV\ lines observed with the
\acl{IRIS} (IRIS, \citeads{2014SoPh..289.2733D}) 
which commonly are interpreted assuming \acdef{CE}.
This \rev{condition is} the opposite of \acp{SB} equilibrium by
requiring statistical equilibrium in which collisional deexcitation
and recombination are fully negligible instead of fully dominating.

A corollary of \PubI\ is that other \Halpha\ features with hot and
dense onsets may similarly gain large \acp{SB} extinction at such
moments and then retain these high values during \rev{subsequent cooling}.
Reversely, features with unusual opaqueness in \Halpha\ may represent
\rev{products} of hot and dense onset phenomena. 
\rev{Since} long fibrils represent a class of \Halpha\ features with
intriguing opacities, we \rev{wondered} whether preceding hot
instances might be found \rev{producing them}. \revpar
The question became how to identify such \revpar precursors.

A 2014 multi-telescope campaign \rev{utilizing} ``last photons''
\rev{of} the SUMER spectrometer
(\citeads{1995SoPh..162..189W}) 
\rev{was undertaken} \rev{to search for} hot precursors in Lyman
lines, \revpar but \rev{without success}.
\revpar \rev{We then} found a striking example \rev{in} \revpar
high-resolution \Halpha\ imaging spectroscopy from the \acl{SST} (SST,
\citeads{2003SPIE.4853..341S}) 
together with hotter diagnostics from the \acl{AIA} (AIA,
\citeads{2012SoPh..275...17L}) 
onboard the \acl{SDO} (SDO,
\citeads{2012SoPh..275....3P}), 
and \revpar saw it confirmed in \revpar \rev{sharper} images from \acp{IRIS}. 
\rev{Here} we present and discuss this example, which we call
``contrail fibril''. \revpar

The observations are detailed in the next section.
The displays combine extracts of the multiple data sets that, when
viewed and blinked as multi-diagnostic movies, led to our noticing the
feature.
In Sect.~\ref{sec:interpretation} we estimate the \rev{visibilities}
of \rev{its} onset, \rev{contrail,} and aftermath. \revpar In the
discussion (Sect.~\ref{sec:discussion}) we argue that the phenomenon
shared characteristics with spicules-II and speculate about why such
features \rev{become} so long, \revpar what they \revpar tell us about
magnetic field topography, \rev{and how ubiquitous they are}.
\revpar We conclude the \rev{study} and outline follow-up in
Sect.~\ref{sec:conclusion}. 

%% fig:haprofs
%===========================================================================
\begin{figure}
  \centerline{\includegraphics[width=0.95\columnwidth]{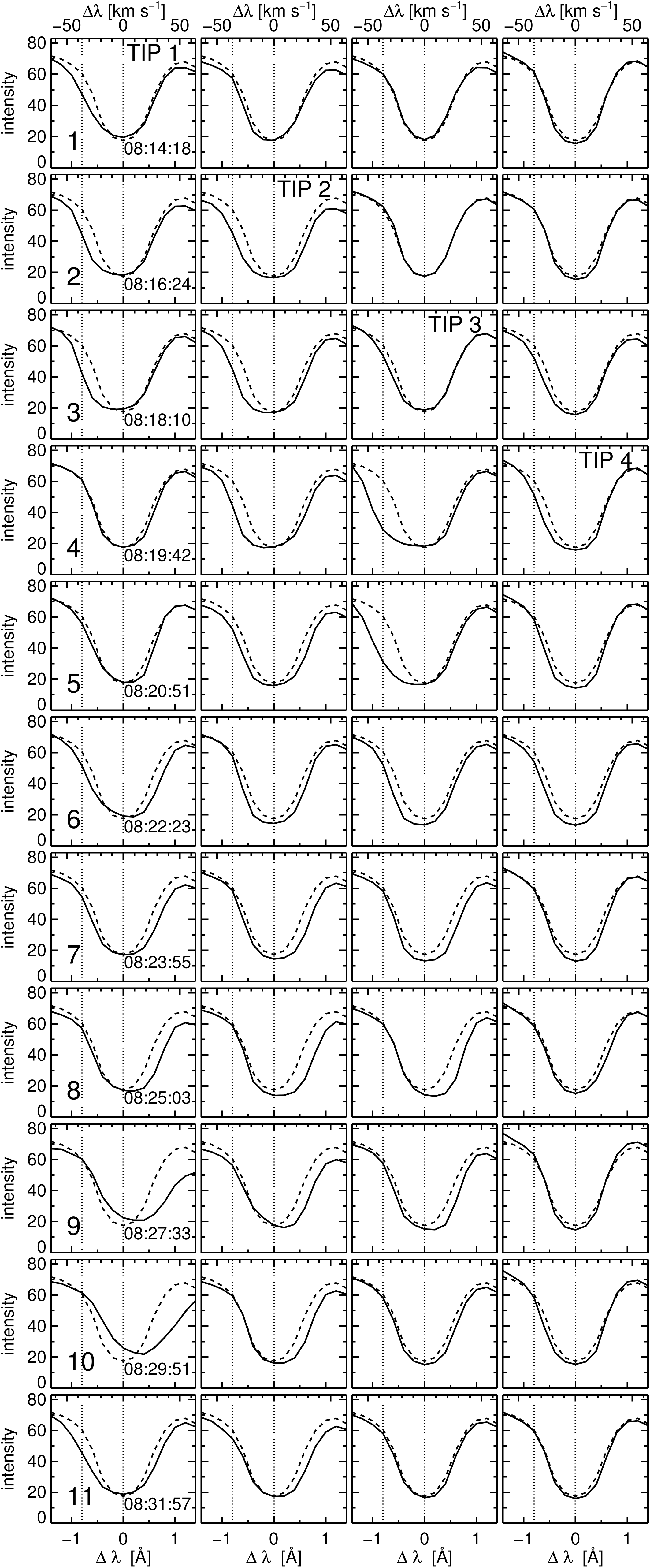}}
  \caption[]{\label{fig:haprofs} 
  \Halpha\ profiles at the four locations marked by arrows in
  Fig.~\ref{fig:sequence}.  
  The wavelength separation from line center is given in \AA\ at the
  bottom, in \kms\ at the top.
  The intensity scale is arbitrary. 
  The first nine rows are the same time samplings and have the same
  numbers as in Fig.~\ref{fig:sequence}, similarly specified in the
  first column. 
  The last four rows are for the same times as in
  Fig.~\ref{fig:aftermath}.
  The columns correspond to the initial precursor-tip locations marked
  by the four arrows in Fig.~\ref{fig:sequence}. 
  The panels for their defining moment are labeled ``TIP'' in the
  first four rows.
  {\em Solid\/}: \Halpha\ profile at this location and time.
  {\em Dashed\/}: mean profile of the whole sequence for reference,
  identical in all panels. 
  {\em Dotted, vertical\/}: sampling wavelengths used in
  Fig.~\ref{fig:sequence}.
  }
\end{figure}
%===========================================================================

%%%%%%%%%%%%%%%%%%%%%%%%%%%%%%%%%%%%%%%%%%%%%%%%%%%%%%%%%%%%%%%%%%%%%%%%%%%%
\section{Observations} \label{sec:observations}
%%%%%%%%%%%%%%%%%%%%%%%%%%%%%%%%%%%%%%%%%%%%%%%%%%%%%%%%%%%%%%%%%%%%%%%%%%%%

\paragraph{Data collection and reduction.}
%%%%%%%%%%%%%%%%%%%%%%%%%%%%%%%%%%%%%%
For this study we analyzed data from a joint \acp{SST}--\acp{IRIS}
observing campaign targeting a quiet area near solar disk center.
Figure~\ref{fig:full} gives an overview.  

On June 21, 2014 the second author obtained imaging spectroscopy at
the \acp{SST} with the \acl{CRISP} (CRISP,
\citeads{2008ApJ...689L..69S}) 
during 08:02--09:15~UT at a cadence of 11.5\,s.
\Halpha\ was sampled at 15 wavelengths spanning $\pm 1.4$\,\AA\ from
line center, \CaIR\ at 25~wavelengths spanning $\pm 1.2$\,\AA\ from
line center, Stokes-$V$ magnetograms were obtained in \FeI\,6303\,\AA.
The reduction used the CRISPRED pipeline of
\citetads{2015A&A...573A..40D}.
It included dark and flat field correction, multi-object multi-frame
blind deconvolution (\citeads{2005SoPh..228..191V}), minimization of
remaining small-scale deformation through cross-correlation
(\citeads{2012A&A...548A.114H}), prefilter transmission correction
following \citetads{2010PhDT.......219D}, correction for
time-dependent image rotation due to the alt-azimuth telescope
configuration, and removal of remaining rubber-sheet distortions by
destretching (\citeads{1994ApJ...430..413S}). 

The same region was observed with \acp{IRIS}. 
Unfortunately, its scanning spectrograph slit missed the feature
described here so that we only use slitjaw images in the 1400\,\AA\
passband. 
These are dominated by the \SiIV\ doublet at 1394\,\AA\ and 1403\,\AA\
in high-temperature gas (\eg\
\citeads{2015ApJ...812...11V}). 

In addition, we collected corresponding image sequences from
\acp{SDO}.
All image sequences were rotated to the \acp{SST} image orientation
and precisely co-aligned using interpolation of the space-based images
to the spatial and temporal grid of the \acp{SST} images.
This was done with the \wlsolarsoft{SolarSoft} library and \acp{IDL} programs
available on the \wlRRtop{website of the first author}. 

The \acp{PSBE} recipe defined in \PubI\ led us to search for features with
large \Halpha\ extinction that follow on hot-onset precursors.
While inspecting the \acp{SST} and \acp{AIA} diagnostics using the
\acl{CRISPEX} (CRISPEX; \citeads{2012ApJ...750...22V}) as browser 
to compare and blink time-delay movies, we quickly noted that the
prominent dark \Halpha\ fibril in Fig.~\ref{fig:full} was preceded
minutes earlier by a fast-moving bright blob in \HeII\,304\,\AA\ and
that the \Halpha\ fibril outlined the passage of this disturbance.
Subsequently, we found that the latter was mapped very well in
the outer blue wing of \Halpha.
When we then co-aligned the 1400\,\AA\ slitjaw images from \acp{IRIS}
we found brightening that clearly tracks the same disturbance.  
\rev{Figures~\ref{fig:full}--\ref{fig:haca}} present the phenomenon.

\paragraph{Field of view.}
%%%%%%%%%%%%%%%%%%%%%%%%
Figure~\ref{fig:full} shows the observed region. 
It was \revpar very quiet, \revpar far from AR\,12093 and AR\,12094
on the Southern hemisphere. 
The first panel shows that it contained only a small pore besides
granulation and tiny magnetic network concentrations in intergranular
lanes.
At the \acp{SST} resolution the latter appear as bright points even in
the continuum (we invite the reader to zoom in per pdf viewer).
The magnetogram in the second panel shows that these together
constituted fairly dense patches of active network. 
The other diagnostics show corresponding brightness features.

The blue wing of \CaIR\ (third panel) maps \revpar magnetic concentrations
as enhanced brightness and elsewhere displays reverse
granulation in the mid photosphere
(\citeads{2011A&A...531A..17R}). 
The \Halpha\ core samples the fibrilar chromosphere.
In this quiet field long fibrils extend from the network patches but
are less ubiquitous elsewhere. 
Some seem to connect the main opposite-polarity patches across the
center of the field, but our feature, the large black fibril
within the overlaid frame, does not \rev{reach that far}.
Elsewhere, notably in the lower-left corner, the \Halpha\ image shows
non-fibrilar swirly brightness patterns which \rev{probably} represent
acoustic shock patterns
(\citeads{2008SoPh..251..533R}). 

The diagnostics in the lower row show increasingly diffuse brightness
response to the enhanced network areas.
The 1400\,\AA\ panel shows \rev{largest brightness }near the polarity
\rev{reversal} between the main field concentrations, possibly from
\rev{earlier} reconnection, and elsewhere bright points as the ones
ascribed to acoustic shocks by
\citetads{2015ApJ...803...44M}. 

The three \acp{AIA} panels show a larger brightness blob above the
network patches indicating heating of the higher atmosphere.
\rev{There are} similarities, even in fine structure, between \HeII\,304 and
\FeII\,193\,\AA\ and \revpar more diffuse veiling originating further
to the left that seems to cover this fine structure in
\FeII\,193\,\AA\ and yet more in \FeIX\,171\,\AA.

\paragraph{Precursor and contrail.}
%%%%%%%%%%%%%%%%%%%%%%%%%%%%%%%%
Figure~\ref{fig:sequence} details the dark \Halpha\ line-center fibril
that we call the 
``contrail fibril'' and is the topic of this study. 
It is prominent in the second column in rows 5 -- 8, longest in row 6.
It is much weaker or absent in the preceding rows, but the blue-wing
images in the first column show a long, slender, growing dark feature
at the same location \rev{with bright counterparts in the
\acp{IRIS} and \acp{AIA} columns}.

The action started with a fan of \acp{RBE} extensions near the foot of
the contrail fibril.
It first shot off a long \acp{RBE} towards the upper left, sampled in
the first panel of Fig.~\ref{fig:sequence}. 
Unfortunately, the \acp{IRIS} slit had just passed (dark slanted line
in the \rev{top} 1400\,\AA\ panel). 

Subsequently, the dark \revpar \rev{extensions} seen in the second
\Halpha-wing panel (we again invite the reader to zoom in per pdf
viewer) \rev{fanned out} progressively in shooting off \rev{growing}
\acp{RBE}s from left to right \rev{as in a peacock tail display}, finally
\rev{sending} off the long precursor of our contrail fibril towards
the right. 
\rev{The latter's extension progress is marked by the four arrows. 
Each was placed successively at the tip of the growing dark blue-wing
feature in rows 1--4. 
The \Halpha\ profiles in Fig.~\ref{fig:haprofs} are for these four
locations.

The second arrow actually identifies the tip of another \acp{RBE} shot
off in parallel to the main one pointed at later by arrows 3 and 4. 
This was also a precursor: it later produced its own, shorter, and
more curved contrail (dark fibril in \Halpha\ line center in rows
4--6). 
We call this contrail B.

The mean proper motion speeds over the surface defined by the
arrow-marked tip locations of the main dark streak in the initial
blue-wing images in the first column of Fig.~\ref{fig:sequence} are
13, 15, and 53~\kms\ for 1$\rightarrow$2, 2$\rightarrow$3, and
3$\rightarrow$4, respectively.
It accelerated appreciably. 
The last value is similar to the speeds reached by \acp{RBE}s 
(\citeads{2009ApJ...705..272R}), 
but this streak was much longer when it} reached 
\rev{15-arcsec} extension \rev{five} minutes \rev{after its launch} (row
4; \rev{observing times are specified in the first column)}.

\rev{The dark fibril at the center of \Halpha\ appeared a minute
later and lasted four minutes (second column, rows 5--8).
It faithfully mapped the precursor shape and therefore represents a
post-passage contrail.}

Actually, we first noted the extending precursor in the \HeII\
304\,\AA\ sequence (fourth column) where it is most clearly evident in
rows 3--6.
We then recognized its presence (when you know what to look for) also
in the \FeIX\ 171\,\AA\ and \FeXII\ 193\,\AA\ images.
We sample only the latter in Fig.~\ref{fig:sequence} in order not to
make its panels too small, but \rev{Fig.~\ref{fig:haca} shows the
corresponding \FeIX\ 171\,\AA\ cutouts at the times of rows 4 and 6 of
Fig.~\ref{fig:sequence}.}

Only afterwards we noted the slender precursor in the \Halpha\ wing
images and realized from inspecting the \Halpha\ profiles with
\acp{CRISPEX} that it is so dark from \rev{\acp{RBE} signature
including large} line broadening.
\rev{This is detailed with profile samples in Fig.~\ref{fig:haprofs}.}
\revpar

Later we co-aligned the \acp{IRIS} images and found that the precursor
is very evident in the 1400\,\AA\ slitjaw images (third column of
Fig.~\ref{fig:sequence}). 
\rev{In rows 3--5 it appears as a very thin extending streak starting
at location 3 and extending to but not reaching location 4.
An underlying bright point, also present before and after, was
enhanced by the streak which implies optically thin formation of the
\SiIV\ lines in the streak and therefore direct imaging of intrinsic
fine structure, in particular the narrow precursor width.}

Figure~\ref{fig:sequence} shows that the precursor became hotter along
its track.
\rev{The 1400\,\AA\ and hotter diagnostics do not show its start from
location 1 to location 3 in rows 1--3, only the part from 3 to 4 in
rows 4--6, 
but in 1400\,\AA\ the streak does not reach location 4.
In \HeII\ 304\,\AA\ and \FeXII\ 193\,\AA\ the streak is most evident
closer to location 4 (row 4), and well-evident in
\FeIX\,171\,\AA\ between these locations (Fig.~\ref{fig:haca}).
This suggests increasing ionization along the track to very high
degree, with \SiIV\ peaking closer to location 3 and the highest iron
stage closest to location 4.}

\revpar \rev{In summary, a} sudden event in the \rev{low} atmosphere
sent off \rev{a} disturbance that heated gas along its \rev{path} to
\rev{very} high temperature.

By row 6, seven minutes after its onset, the precursor \revpar
\rev{became invisible}, but by then had left the black fibril
at \Halpha\ line center \rev{along its wake}.
\rev{This contrail persisted} three more minutes. 

By 08:27\,UT (row 9) the show was over \rev{in these high-temperature
diagnostics} -- just when the \acp{IRIS} slit was nearing in the next
scan (\rev{bottom} 1400\,\AA\ panel). \revpar \rev{It is a pity that
the scan timing missed the \SiIV\ emission streak in rows 4--6 because
the slit happened to be aligned with its direction.}

%% fig:aftermath
%===========================================================================
\begin{figure*}
  \sidecaption
  \includegraphics[width=12cm]{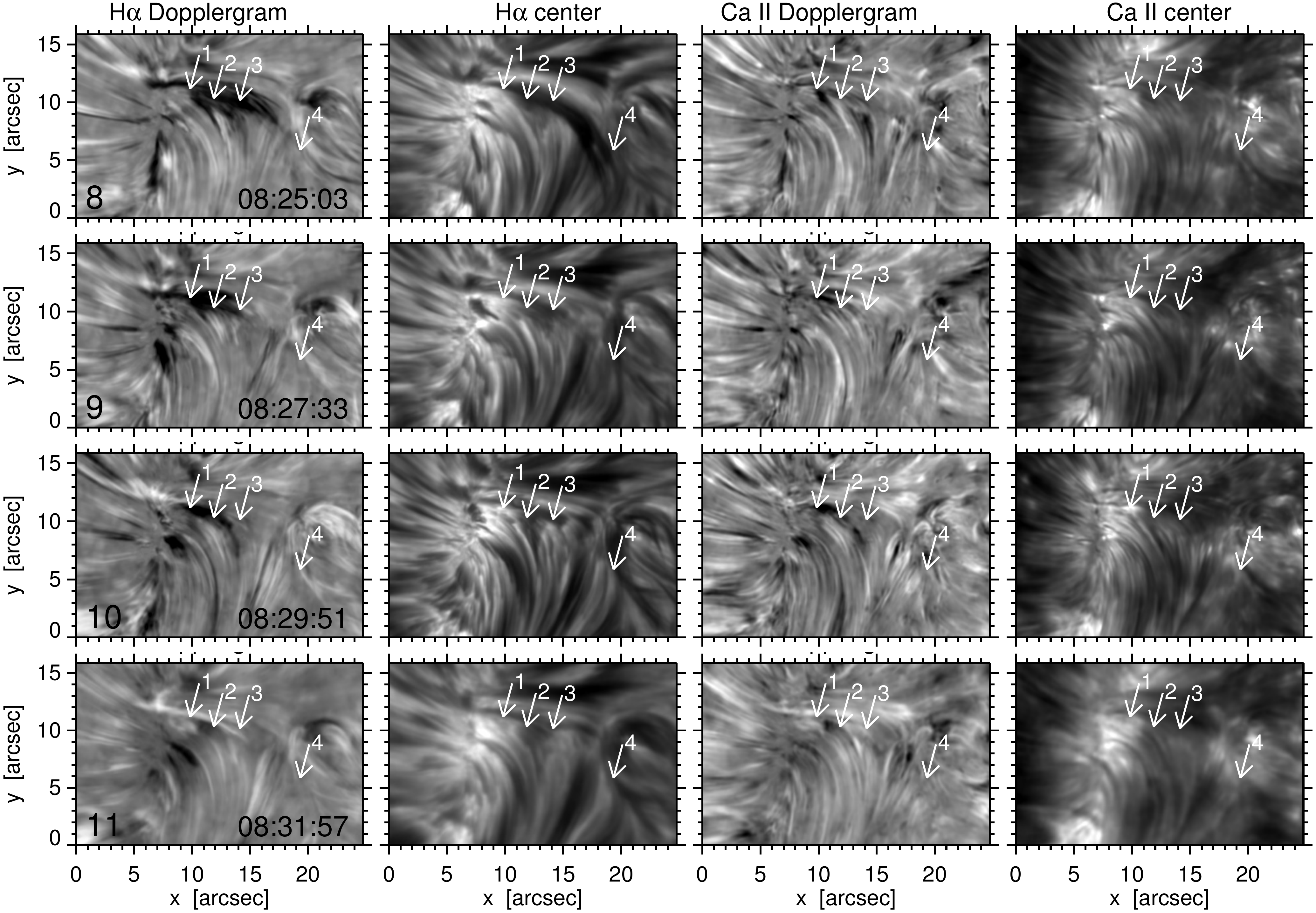}
  \caption[]{\label{fig:aftermath} 
  Time sequence of aftermath image cutouts as in
  Fig.~\ref{fig:sequence}.
  {\em Columns\/}: \Halpha\ Dopplergrams at
  $\Delta \lambda \tis \pm 0.6$~\AA\ with black denoting redshift.
  \Halpha\ images at line center, \CaIR\ Dopplergrams at
  $\Delta \lambda \tis \pm 0.3$~\AA, \CaIR\ images at line center.
  The first two rows are for the same moments as the last two rows in
  Fig.~\ref{fig:sequence} and share their numbers.
  The intensity images are bytescaled individually but the
  Dopplergrams have identical greyscales along columns.
  \vspace{2ex}\mbox{} 
  }
\end{figure*}
%===========================================================================

\rev{
\paragraph{Comparison with Ca\,II\,8542\,\AA.}
%%%%%%%%%%%%%%%%%%%%%%%%%%%%%%%%%%%%%%%%
Figure~\ref{fig:haca} adds cutout images in \CaIR\ and
\FeIX\,171\,\AA\ for key moments in Fig.~\ref{fig:sequence}: the time
when the precursor streak reached its maximum length in the blue wing
of \Halpha\ (row 4) and nearly three minutes later (row 6) when the
line-center contrail fibril reached maximum visibility.
The \CaIR\ wing image shows the first part of the precursor streak,
but not beyond location 3. 
The contrail fibril is not evident at all in \CaIR.

\paragraph{H-alpha profiles.}
%%%%%%%%%%%%%%%%%%%%%%%%%%
Figure~\ref{fig:haprofs} shows \Halpha\ profiles for the locations
defined by the four arrows in Fig.~\ref{fig:sequence} and at the same
sampling times, so with the same row numbers and time labels as in
Fig.~\ref{fig:sequence}. 
The four panels labeled TIP in the first four rows correspond to the
location defining times in Fig.~\ref{fig:sequence} marking how far the
extending precursor streak had reached in the \Halpha\ blue-wing images. 
Each tip profile shows characteristic \acp{RBE} shape with
substantial core broadening on the blue side.
This signature increased at all four locations for some minutes after
the tip definition, in particular at location 3.   

Generally, \Halpha\ broadening suggests temperature increase with less
influence from nonthermal line broadening than for other lines due to
the small atomic weight of hydrogen
(\citeads{2009A&A...503..577C}). 
Such profiles therefore indicate heating plus updraft.

The \acp{RBE} signature vanished by row 6, leaving just hotter and
darker profiles than average when the precursor lost its visibility in
the blue wing of \Halpha\ and in the \acp{IRIS} and \acp{SDO} images. 
The heating event had passed and left cooling gas with large \Halpha\
opacity producing the contrail fibril at \Halpha\ center (rows 5--8 of
Fig.~\ref{fig:sequence}).

From row 7 most profiles show reversed Dopplershift signatures detailed
in the next paragraph.

\paragraph{Cooling aftermath.}
%%%%%%%%%%%%%%%%%%%%%%%%%%%%
Rows 7--11 of Fig.~\ref{fig:haprofs} show first increasing and then
decreasing high-temperature-plus-redshift signature for locations
1--3. 
In rows 7 and 8 it was largest at location 2 but in rows 9--10 it
became rather outrageous at location 1.
This suggests an aftermath in which kicked-up heated gas moved back
down while recombining and becoming invisible in the hotter
diagnostics.
Since hydrogen recombination takes much energy the gas cooled
drastically and at about 10\% neutral passed into the regime with
retarded hydrogen recombination below 7000\,K as in the fundamental 1D
non-equilibrium simulation of
\citetads{2002ApJ...572..626C} 
and in the post-shock gas in the 2D HION atmosphere. 
During this retarded phase the cooling gas maintained large \Halpha\
hot-precursor opacity.
This is confirmed by the cutout images in Fig.~\ref{fig:aftermath}
which show a retracting feature with large redshift. 
Only its base between locations 1 and 2 became visible also in the
\CaIR\ Dopplergrams, best in row 10, at smaller opacity.

Both line centers brightened at location 1 in rows 9 and 10 of
Fig.~\ref{fig:aftermath} from the Dopplershift evident in the first
column of Fig.~\ref{fig:haprofs}.

In the final row of Fig.~\ref{fig:aftermath} the gas at location 1 was
still unusually hot (Fig.~\ref{fig:haprofs}), but already showed
blueshift due to a new heating agent drawing a thin bright streak from
left to right in the Dopplergrams. 
It just missed the other sampling locations.
This streak became visible also in the \acp{IRIS} 1400\,\AA\ images,
extended to 15\,arcsec length, and two minutes later produced a long
\Halpha\ line-center fibril much as the one presented which
was also followed by a contracting redshift feature.
It was very similar to our contrail except slenderer. 
We call it contrail C.
}

\paragraph{Conclusion of this section.}
%%%%%%%%%%%%%%%%%%%%%%%%%%%%%%%%%%
\revpar If one had noted the dark \rev{\Halpha\ } fibril when it was
most prominent, in rows 6--8 of Fig.~\ref{fig:sequence}, and tried to
model that (say with cloud modeling) without being aware of the hot
precursor in rows 3--5, \rev{then} one would be as mistaken as when
interpreting jet contrails in the sky without knowing about \rev{jet
engines}.
\rev{It was an intrinsically dynamic phenomenon starting with a fierce
jet-like heating agent producing gas at very high temperature which
subsequently recombined and cooled, became visible at the center of
\Halpha, and partly flowed back. 
Also a jet contrail, but from a solar jet and recombination rather
than condensation.}

%%%%%%%%%%%%%%%%%%%%%%%%%%%%%%%%%%%%%%%%%%%%%%%%%%%%%%%%%%%%%%%%%%%%%%%%%%%%
\section{Interpretation}    \label{sec:interpretation}
%%%%%%%%%%%%%%%%%%%%%%%%%%%%%%%%%%%%%%%%%%%%%%%%%%%%%%%%%%%%%%%%%%%%%%%%%%%%
\rev{In this section we do not try to interpret the nature and
mechanism of the heating event but address the visibilities of the
precursor, contrail, and aftermath.
}

\paragraph{PSBE  estimation.}
%%%%%%%%%%%%%%%%%%%%%%%%%%
One would not expect to see the same feature in \Halpha, \HeII, \SiIV,
\FeIX\ and \FeXII\ simultaneously, but in rows 4 and 5 of
Fig.~\ref{fig:sequence} the onset streaks correspond fairly closely
between these diagnostics.
Obviously, they are due to some process causing large heating \revpar
(Sect.~\ref{sec:discussion}). \revpar \rev{It was initiated} \revpar
\rev{along the row of dark extensions
in the \Halpha\ blue-wing image in row 2 of Fig.~\ref{fig:sequence}}
in the low atmosphere, \revpar implying dense gas for the onset. 
\revpar The combination of hot and dense makes the \rev{\acp{PSBE}}
recipe of \PubI\ \revpar
\rev{appropriate} for \revpar explaining the joint \rev{precursor}
visibilities in these very disparate diagnostics \rev{ and the
subsequent appearance of the \Halpha\ line-center contrail and its
aftermath.}

%% ============ figures interpretation

%% fig:CE-LTE
%===========================================================================
\begin{figure}
  \centerline{\includegraphics[width=\columnwidth]{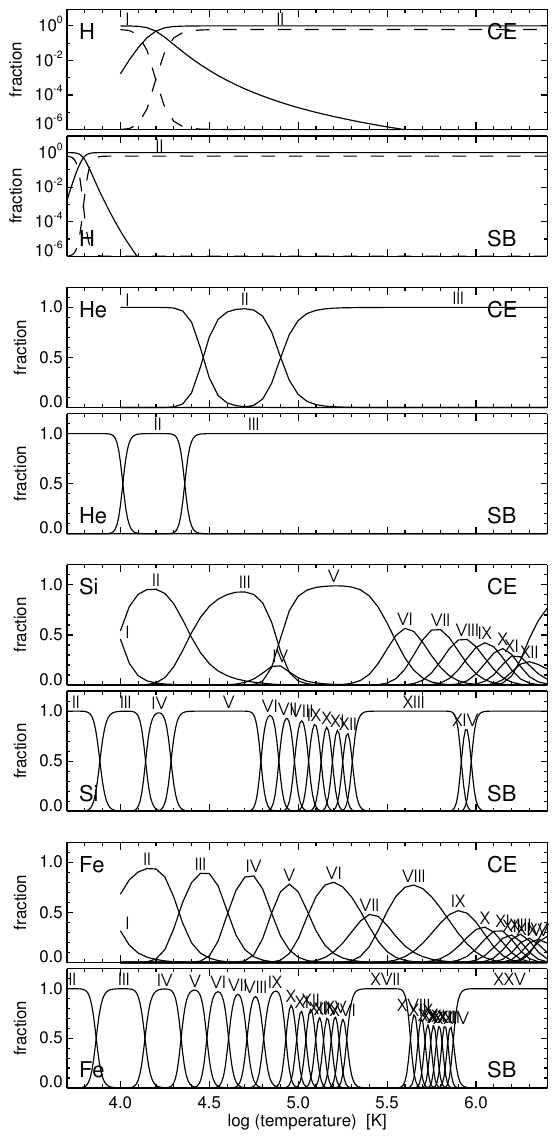}}
    \caption[]{\label{fig:CE-LTE} 
    \acp{CE}/\acp{SB} ionization-stage population comparisons for H,
    He, Si and Fe.
    {\em Upper panel of each pair\/}: \acp{CE} distribution with
    temperature.
    {\em Lower panel of each pair\/}: \acp{SB} distribution with
    temperature for fixed electron density
    $N_\rme \tis 10^{14}$~cm$^{-3}$.
    For lower $N_\rme$ the \acp{SB} curve patterns remain
    similar but the flanks steepen and the peaks shift leftward
    (about $-0.05$ in $\log(T)$ for tenfold $N_\rme$ reduction).
    The first pair for hydrogen has logarithmic $y$-axes; the dashed
    curves are on the linear scales of the other panels.  
    }
\end{figure}
%===========================================================================

%% fig:extLTE
%===========================================================================
\begin{figure}
  \centerline{\includegraphics[width=0.8\columnwidth]{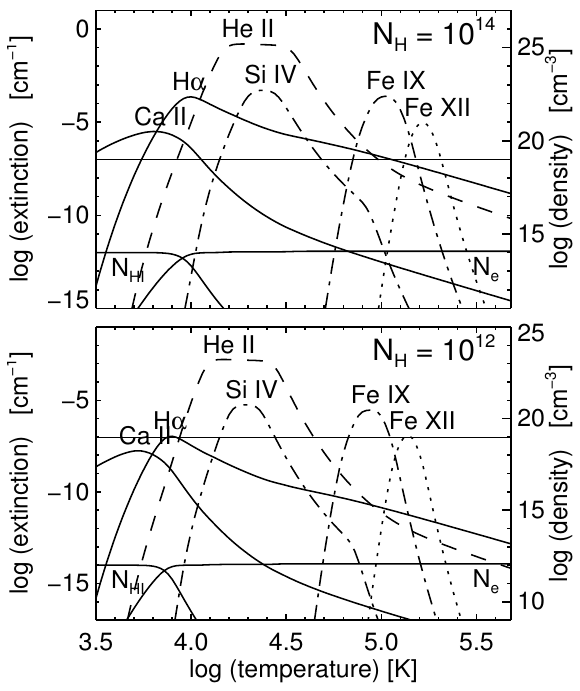}}
    \caption[]{\label{fig:extLTE} 
    \acp{SB} extinction $\alpha^l$ for \revpar \Halpha\
    (solid), \CaIR\ (solid), \HeII\,304\,\AA\ (dashed),
    \SiIV\,1394\,\AA\ (dot-dashed), \FeIX\,171\,\AA\ (dot-dashed) and
    \FeXII\ 193\,\AA\ (dotted) as function of temperature, for total
    hydrogen densities \rev{$N_\rmH\tis10^{14}$ and
    $10^{12}$~cm$^{-3}$ (upper and lower panel, respectively)}. 
    The two cross-over solid curves near the bottom are the
    \rev{competing} neutral hydrogen and electron density,
    with scales on the right.
    The extinction scales at left and the density scales at right
    shift upward between \rev{the} panels to compensate for the density
    decrease.
    The horizontal line at $\log \alpha^l \tis-7$ marks optical
    thickness unity for a slab of 100~km geometrical thickness.
    \revpar
    }
\end{figure}
%===========================================================================

\rev{The recipe employs} Figs.~\ref{fig:CE-LTE} and \ref{fig:extLTE}.
They resemble Figs.~4 and 5 of \PubI\ and were made as described
there.
Figure~\ref{fig:CE-LTE} compares \acp{CE} ionization distributions
with their \acp{SB} counterparts over a wide temperature range.
Figure~\ref{fig:extLTE} quantifies \acp{SB} extinction coefficients
for the specified lines as function of temperature, for \revpar
total hydrogen densities \revpar corresponding to heights \revpar
850 and 1520\,km \revpar in \rev{the chromosphere of} the ALC7 standard
model of \citetads{2008ApJS..175..229A}. 

\revpar

\revpar \rev{Figure~\ref{fig:CE-LTE} quantifies the large
differences in the temperatures at which successive ionization stages
of the plotted elements reach maximum presence between the \acp{SB}
and \acp{CE} extremes.
For example, the \acp{IRIS} \SiIV\ lines typically form at 80\,000\,K
in \acp{CE} but at 16\,000\,K when \acp{SB} holds. 
Such pair comparison defines the range from \acp{SB} validity at very
high density to \acp{CE} validity at very low density.}

\acp{SB} extinction \rev{can} be valid in momentary hot and dense
instances \rev{at chromospheric heights where generally \acp{NLTE} is
the rule}.
When the gas becomes so hot that hydrogen ionizes substantially, the
resulting large boost of the electron density may create collisionally
constrained \acp{SB} populations at sufficiently high gas density
\rev{and feature thickness}. 
This then holds {usually only} for lower levels of resonance lines,
not for their upper levels because these suffer \acp{NLTE} photon
losses. \revpar

\rev{\Halpha\ is a special case because it rides on top of \Lyalpha\
in which even a small} neutral-hydrogen feature can be sufficiently
opaque to thermalize \revpar \rev{\Lyalpha} radiation \revpar \rev{
notwithstanding the very} small \rev{\Lyalpha} collision \rev{rates}.
This is the case throughout the ALC7 chromosphere (Fig.~1 of \PubI)
and also within the HION shocks.
\rev{In the latter} the \Halpha\ extinction \revpar reaches the
\acp{SB} value at electron densities as low as $10^9$\,cm$^{-3}$
thanks to 10\% hydrogen ionization producing $10^3$ more electrons
than available from the electron-donor metals in cooler gas (Fig.~2 of
\citeads{2007A&A...473..625L}). 
\revpar

\rev{\acp{SB} extinction is also a good assumption for \CaIR\ (\PubI)
and applies to resonance lines of majority species wherever the Saha
law applies, which at high density is more likely than \acp{CE}.}
Therefore, \revpar the curves \revpar \rev{for \Halpha\ and \CaIR\ in
Fig.~\ref{fig:extLTE} are probably correct where they peak, whereas
the temperature locations of the peaks for the higher ions represent 
lower limits only.
\revpar However,} even for these \rev{peaks} shifts as far to the right as
predicted by \acp{CE} (Fig.~\ref{fig:CE-LTE}) are unlikely.

\paragraph{Precursor  visibilities.}
%%%%%%%%%%%%%%%%%%%%%%%%%%%%%%
The \Halpha\ curves in Fig.~\ref{fig:extLTE} are for line center. 
For the \Halpha\ wing visibility they should be shifted down over
about one logarithmic unit
and so get significantly below the horizontal line \revpar in the
\rev{lower} panel. \revpar The choice of a 100-km slab as geometrical
feature thickness is \rev{arbitrary}. \revpar The width of the
blue-wing precursor in the first column of Fig.~\ref{fig:sequence}
measures about 1\,arcsec or 700\,km \rev{but this is widened by
resonance scattering in \Lyalpha\ (\PubI).
The intrinsic precursor width shown by the 1400\,\AA\ panels is
sub-arcsecond. 
The 100-km line should therefore be a reasonable visibility
indicator.}
\revpar

Because the \Halpha\ curves drop less steeply then the others 
\revpar \Halpha\ may still reach overlap visibility with \HeII\ and
\SiIV\ as \rev{predicted in the upper panel of Fig.~\ref{fig:CE-LTE}
and} observed in the first, third and fourth panels of row 4 of
Fig.~\ref{fig:sequence} \rev{(remember that both extinction and
emissivity scale with the lower-level population)}.
Since the initial part of the \Halpha-wing precursor is not visible in
\rev{these hotter diagnostics} it started \rev{well} below 15\,000\,K
but then became much hotter.

\rev{The \CaIR\ visibility constrains the initial temperature of the
precursor to yet lower values. 
The curves for this line and for \Halpha\ in Fig.~\ref{fig:extLTE}
cross at $T \tapprox 6000$\,K with nearly opposite temperature
sensitivity.
The precursor start near location 1 shows similar morphology at the
two line centers in Fig.~\ref{fig:haca}, suggesting temperature well
below 10\,000\,K, but beyond location 3 the precursor is invisible in
\CaIR, suggesting temperature around or above 10\,000\,K.  The 
subsequent contrail also remained invisible (Figs.~\ref{fig:haca} 
and \ref{fig:aftermath}).}

The \rev{Fe-line} humps in Fig.~\ref{fig:extLTE} are around
$T \tis 10^5$\,K. 
\rev{Since the precursor tip showed up in them it must have become}
very hot, even if it did not become \acp{CE} hot
($T \tapprox 10^6$\,K).
\revpar \rev{The observations in these lines (Figs.~\ref{fig:sequence}
and \ref{fig:haca}) show best precursor visibility near location 4 at
the time of row 4, but the precursor also reached that far in \HeII\
304\,\AA\ and in the blue wing of \Halpha.
This joint visibility suggests temperature around $10^5$\,K at
hydrogen density around $10^{14}$~cm$^{-3}$, yet hotter for conditions
closer to \acp{CE}.}

\revpar

\paragraph{Contrail and aftermath visibilities.}
%%%%%%%%%%%%%%%%%%%%%%%%%%%%%%%%%%%%%%%%
The \rev{\acp{PSBE} recipe} \revpar also explains the appearance of
the dark fibril at \Halpha\ line center well after the \revpar passage
of the heating event.   
The latter ionized hydrogen \rev{instantaneously} but hydrogen
recombination in the \rev{subsequent} cooling gas was retarded. 
\revpar While the gas cooled to \rev{below $10^4$\,K} the \Halpha\
extinction did not slide down the steep \acp{SB} \rev{slopes} in
Fig.~\ref{fig:extLTE} instantaneously, but instead stayed \rev{near
the peak value passed at about 7000\,K} during multiple
minutes. \revpar \rev{Figure~\ref{fig:haprofs} shows that this
relaxation happened first at location 4 and progressively later at the
lower-number locations along which the heated gas came back down
during the aftermath.

At location 1 near the base of the heating event in the low atmosphere
the cooling was largest and took the longest.
There the temperature lowered sufficiently for \CaIR\ Dopplershift
visibility (Fig.~\ref{fig:aftermath}). 
The \CaIR\ curves in Fig.~\ref{fig:extLTE} peak near 5000\,K; at such
temperatures the instantaneous \Halpha\ extinction is very much lower,
but retardation kept the actual \Halpha\ opacity higher as evident
in Fig.~\ref{fig:aftermath}.}

\paragraph{Conclusion of this section.}
%%%%%%%%%%%%%%%%%%%%%%%%%%%%%%%%%%%%
\rev{The \acp{PSBE} recipe applies well to the contrail phenomenon. 
It explains that the precursor was observed simultaneously in the very
disparate diagnostics in Fig.~\ref{fig:sequence} but only its start in
\CaIR, that a large dark subsequent contrail fibril followed in the
center of \Halpha, and that the cooling returning aftermath cloud
retained large \Halpha\ opacity while eventually becoming visible 
in \CaIR.
}

%%%%%%%%%%%%%%%%%%%%%%%%%%%%%%%%%%%%%%%%%%%%%%%%%%%%%%%%%%%%%%%%%%%%%%%%%%%%
\section{Discussion}\label{sec:discussion}
%%%%%%%%%%%%%%%%%%%%%%%%%%%%%%%%%%%%%%%%%%%%%%%%%%%%%%%%%%%%%%%%%%%%%%%%%%%%

\paragraph{Nature of the event.}
%%%%%%%%%%%%%%%%%%%%%%%%%%%%
The onset \rev{was} much like the start of a regular \acp{RBE}.
Indeed, the detection criteria of
\citetads{2009ApJ...705..272R} 
would have classified this feature as an \acp{RBE}.
The main difference is that its track is so long and that it later got
outlined by the subsequent contrail fibril over such long length.
We speculate that the precursor was essentially a spicule-II
phenomenon but directed more horizontally than in \rev{regular} \acp{RBE}s and
\acp{RRE}s \rev{and possibly from a fiercer heating agent}. 
It originated in a similarly quiet area as in
\citetads{2009ApJ...705..272R} 
\rev{and speeded up and}
\revpar got as hot as spicules-II \rev{do} on their way up,
\rev{while its lower part returned  similarly to the surface 
(\cf\ \citeads{2011Sci...331...55D}; 
 \citeads{2014ApJ...792L..15P})}. 

The mechanism that shoots off spicules-II and their on-disk \acp{RBE}
and \acp{RRE} counterparts \rev{remains un}known
(\citeads{2012ApJ...759...18P}). 
Since they often originate in unipolar network, direct bipolar
strong-field reconnection \rev{seems} unlikely. 
\revpar
%% (and would produce Ellerman
%% bombs but these occur only in complex active regions).
The peacock fan \rev{sequence that sent} off our
contrail fibril \rev{and contrail B} does suggest reconnection,
perhaps component or fly-by reconnection \rev{as in}
\citetads{2012SoPh..278..149M}. 
\rev{It resembles the fan of peacock jets} \revpar in an umbral
light bridge \rev{reported} by
\citetads{2016A&A...590A..57R}, 
but that was probably bipolar reconnection.

The other driver option is local generation of Alfv\'enic waves.
Spicules-II combine jet production with Alfv\'enic swaying and torsion
modes (\eg\
\citeads{2012ApJ...752L..12D}; 
\citeads{2013ApJ...769...44S}); 
the onset of our feature likely harbored the same.
We speculate that its very long length was contributed particularly by
a vorticity kick since torsion waves are uncompressive and can travel
far before dissipation by mode coupling.

\revpar There is \rev{also} the \rev{issue} whether the precursor was
a bullet-like hot blob or a jet-like heating agent that extended in
length.
The long thin shape of the precursor in Fig.~\ref{fig:sequence}
suggests the latter, but \HeII\ and \SiIV\ have large valence-electron
jumps in their previous ionization stages which may produce retarded
recombination similarly to hydrogen (in which it is caused by the
\Lyalpha\ jump, see
\citeads{2002ApJ...572..626C}; 
for helium retardation see
\citeads{2014ApJ...784...30G}). 
Long tracks \rev{may} then also appear in these diagnostics from cooling gas
after a hot-bullet passage.
\revpar

Finally, the \rev{sequential} visibility of the precursor track, \revpar
contrail fibril, \rev{and aftermath} suggest that the \rev{latter}
consisted of cooling gas in which hydrogen recombined.

\paragraph{Contrail ubiquity.}
%%%%%%%%%%%%%%%%%%%%%%%%%%
An obvious issue is whether the contrail fibril presented here is an
uncommon or a common phenomenon. 
We do not answer this question here because it requires detailed
analysis of multiple datasets, sampling different levels of activity;
we prefer to postpone such larger-volume studies to future reports.
However, naturally we have searched the present dataset for similar
instances and cursorily inspected other \rev{\acp{SST}} datasets as
well. 
The easiest way is to blink \Halpha\ line-center movies with varying
time delay against \Halpha\ blue-wing movies with \acp{CRISPEX}, which
permits time-delay different-wavelength movie \rev{blinking with
simultaneous profile displays}.

The upshot is that, \rev{in addition to contrails B and C which
occurred already within the small cutout field during the short period 
presented here}, we quickly found more
but \rev{also} that not every long \Halpha\ fibril has an easily
identifiable hot precursor. 
We have the impression that it helped much that the present field of
view was very quiet, so that our contrail fibril stood out without
interference from \rev{neighboring others in place and time}.
Areas with denser fibril canopies \revpar present so much
time-dependent confusion that it becomes very difficult to identify
precursors and resulting contrails uniquely and reliably.  

\revpar \revpar

\paragraph{\rev{Contrail} fibrils as field mappers.}
%%%%%%%%%%%%%%%%%%%%%%%%%%%%%%%%%%%%
The contrail fibril did not connect \revpar field from network at one
side of a supergranular cell to network on the other side,
\rev{although it pointed to the opposite-polarity field patch below
the frame in Fig.~\ref{fig:full}. \revpar Short of the latter there
was not enough opposite \revpar polarity for such
short-circuiting.} \revpar

Instead, the fibril delineated cooling gas after the passage of the
onset disturbance; the fibril outlined its path.
Line-tying of plasma charged by hydrogen ionization may well have
aligned the disturbance (whether bullet or jet) initially along field
lines.
\rev{Whether the outlined fields indeed continued to span the
internetwork and closed in the large opposite-polarity patch cannot be
established from the contrail.} 

\rev{Also,} the subsequent fibril outlined azimuthal field direction
as it was \rev{a few} minutes \rev{earlier}.
\rev{It may have suffered} subsequent deformation by flows as when
winds affect \rev{jet} contrails; \rev{indeed, the contrail tip seems
to shift away from location 4 in rows 6--8 of
Fig.~\ref{fig:sequence}.}

The resulting suggestion is twofold: (1) \rev{contrail} \revpar
fibrils do not \rev{visibly} connect opposite polarity fields across
internetwork cells; if they appear to do so this \rev{rather} implies
that the fields from both sides point to each other -- as they do not
only in closed loops but also for unipolar fields that meet and turn
up halfway as in standard cartoons (including the venerable Fig.~1 of
\citeads{1967IAUS...28..293N}), 
and (2) in such two-sided mapping \Halpha\ fibrils do not outline the
present but possibly the past field topography. 
Slender features in the \Halpha\ wings \rev{or \Halpha\ Dopplergrams
and mapped yet better through optically thin line formation} in
1400\,\AA\ images are better instantaneous indicators.

The memory effect adds to the incomplete rendition of field topography
due to the complex interactions of the Alfv\'enic waves that fibrils
harbor, as shown by
\citetads{2015ApJ...802..136L}. 
Their simulation did not include non-equilibrium \Halpha\ synthesis,
therefore lacked the \Halpha\ memory of large preceding hydrogen
ionization, and so underestimated the lack of fibril-field
correspondence.
Proper 3D non-equilibrium spectrum synthesis \rev{remains} too
demanding, but a quick upper estimate is to use the \Halpha\ peaks in
Fig.~\ref{fig:extLTE} by setting the \Halpha\ lower-level population
to $\log (n_2/N_\rmH) = 0.683\,\log(N_\rmH)-14.8$ (\PubI).

%%%%%%%%%%%%%%%%%%%%%%%%%%%%%%%%%%%%%%%%%%%%%%%%%%%%%%%%%%%%%%%%%%%%%%%%%%%%
\section{Conclusion}  \label{sec:conclusion}
%%%%%%%%%%%%%%%%%%%%%%%%%%%%%%%%%%%%%%%%%%%%%%%%%%%%%%%%%%%%%%%%%%%%%%%%%%%%

We have shown how a long dark \Halpha\ fibril appeared along the
trajectory of a fast sudden disturbance which passed minutes before
and heated gas to very high temperature along its track. 
This example shows that \Halpha\ fibrils can represent past happenings
of much fiercer nature than the fibrils themselves would suggest when
interpreted through time-independent modeling of their subsequent
appearance. 
The lesson is that the solar chromosphere is finely structured not
only in space but also in time, and that at least for some features
the recent past must be taken into account to understand their present
presence.

Our obvious next quest is to ascertain whether all or most long
\Halpha\ fibrils are contrails or whether this one was an uncommon
happening. 
We suspect it was not, but that precursor identification is less easy
in fields with larger fibril crowding from more activity.

It will also be good to find contrails sampled by the \acp{IRIS} slit.

%%%%%%%%%%%%%%%%%%%%%%%%%%%%%%%%%%%%%%%%%%%%%%%%%%%%%%%%%%%%%%%%%%%%%%%%%%%%
\begin{acknowledgements}
\rev{We thank Tiago Pereira for his help during the observations.
Comments from the referee led to considerable improvement of the
presentation.}
\acp{IRIS} is a \acp{NASA} small explorer mission developed and
operated by \acp{LMSAL} with mission operations executed at \acp{NASA}
Ames Research Center and major contributions to downlink
communications funded by the Norwegian Space Center through an
\acp{ESA} PRODEX contract.
The \acp{SST} is operated on the island of La Palma by the Institute
for Solar Physics of Stockholm University in the Spanish Observatorio
del Roque de los Muchachos of the Instituto de Astrof{\'\i}sica de
Canarias. 
This research received funding from both the Research Council of
Norway and the European Research Council under the European Union’s
Seventh Framework Programme (FP7/2007- 2013) / ERC grant agreement
nr.\ 291058.
We made much use of the SolarSoft and \acp{ADS} libraries.
\end{acknowledgements}

%%%%%%%%%%%%%%%%%%%%%%%%%%%%%%%%%%%%%%%%%%%%%%%%%%%%%%%%%%%%%%%%%%%%%%%%%%%%
%% references
\bibliographystyle{aa-note}
\bibliography{rjrfiles,adsfiles} 

\begin{thebibliography}{52}
\expandafter\ifx\csname natexlab\endcsname\relax\def\natexlab#1{#1}\fi

\bibitem[{{Avrett} \& {Loeser}(2008)}]{2008ApJS..175..229A}
{Avrett}, E.~H. \& {Loeser}, R. 2008, \apjs, 175, 229 \csname
  2008ApJS..175..229Alink\endcsname~\csname 2008ApJS..175..229Anote\endcsname

\bibitem[{{Beckers}(1964)}]{1964PhDT........83B}
{Beckers}, J.~M. 1964, PhD thesis, Sacramento Peak Observatory, Air Force
  Cambridge Research Laboratories, Mass., USA \csname
  1964PhDT........83Blink\endcsname~\csname 1964PhDT........83Bnote\endcsname

\bibitem[{{Bray} \& {Loughhead}(1974)}]{1974soch.book.....B}
{Bray}, R.~J. \& {Loughhead}, R.~E. 1974, {The solar chromosphere}, ed. R.~J.
  {Bray} \& R.~E. {Loughhead} \csname 1974soch.book.....Blink\endcsname~\csname
  1974soch.book.....Bnote\endcsname

\bibitem[{{Carlsson} \& {Stein}(2002)}]{2002ApJ...572..626C}
{Carlsson}, M. \& {Stein}, R.~F. 2002, \apj, 572, 626 \csname
  2002ApJ...572..626Clink\endcsname~\csname 2002ApJ...572..626Cnote\endcsname

\bibitem[{{Cauzzi} {et~al.}(2009){Cauzzi}, {Reardon}, {Rutten}, {Tritschler},
  \& {Uitenbroek}}]{2009A&A...503..577C}
{Cauzzi}, G., {Reardon}, K., {Rutten}, R.~J., {Tritschler}, A., \&
  {Uitenbroek}, H. 2009, \aap, 503, 577 \csname
  2009A&A...503..577Clink\endcsname~\csname 2009A&A...503..577Cnote\endcsname

\bibitem[{{de la Cruz Rodrigu{\'{e}}z}(2010)}]{2010PhDT.......219D}
{de la Cruz Rodrigu{\'{e}}z}, J. 2010, PhD thesis, Stockholm University \csname
  2010PhDT.......219Dlink\endcsname~\csname 2010PhDT.......219Dnote\endcsname

\bibitem[{{de la Cruz Rodr{\'{\i}}guez} {et~al.}(2015){de la Cruz
  Rodr{\'{\i}}guez}, {L{\"o}fdahl}, {S{\"u}tterlin}, {Hillberg}, \& {Rouppe van
  der Voort}}]{2015A&A...573A..40D}
{de la Cruz Rodr{\'{\i}}guez}, J., {L{\"o}fdahl}, M.~G., {S{\"u}tterlin}, P.,
  {Hillberg}, T., \& {Rouppe van der Voort}, L. 2015, \aap, 573, A40 \csname
  2015A&A...573A..40Dlink\endcsname~\csname 2015A&A...573A..40Dnote\endcsname

\bibitem[{{De Pontieu} {et~al.}(2012){De Pontieu}, {Carlsson}, {Rouppe van der
  Voort}, {Rutten}, {Hansteen}, \& {Watanabe}}]{2012ApJ...752L..12D}
{De Pontieu}, B., {Carlsson}, M., {Rouppe van der Voort}, L.~H.~M., {et~al.}
  2012, \apjl, 752, L12 \csname 2012ApJ...752L..12Dlink\endcsname~\csname
  2012ApJ...752L..12Dnote\endcsname

\bibitem[{{De Pontieu} {et~al.}(2007{\natexlab{a}}){De Pontieu}, {Hansteen},
  {Rouppe van der Voort}, {van Noort}, \& {Carlsson}}]{2007ApJ...655..624D}
{De Pontieu}, B., {Hansteen}, V.~H., {Rouppe van der Voort}, L., {van Noort},
  M., \& {Carlsson}, M. 2007{\natexlab{a}}, \apj, 655, 624 \csname
  2007ApJ...655..624Dlink\endcsname~\csname 2007ApJ...655..624Dnote\endcsname

\bibitem[{{De Pontieu} {et~al.}(2007{\natexlab{b}}){De Pontieu}, {McIntosh},
  {Hansteen}, {Carlsson}, {Schrijver}, {Tarbell}, {Title}, {Shine}, {Suematsu},
  {Tsuneta}, {Katsukawa}, {Ichimoto}, {Shimizu}, \&
  {Nagata}}]{2007PASJ...59S.655D}
{De Pontieu}, B., {McIntosh}, S., {Hansteen}, V.~H., {et~al.}
  2007{\natexlab{b}}, \pasj, 59, S655 \csname
  2007PASJ...59S.655Dlink\endcsname~\csname 2007PASJ...59S.655Dnote\endcsname

\bibitem[{{De Pontieu} {et~al.}(2011){De Pontieu}, {McIntosh}, {Carlsson},
  {Hansteen}, {Tarbell}, {Boerner}, {Mart{\'{\i}}nez-Sykora}, {Schrijver}, \&
  {Title}}]{2011Sci...331...55D}
{De Pontieu}, B., {McIntosh}, S.~W., {Carlsson}, M., {et~al.} 2011, Science,
  331, 55 \csname 2011Sci...331...55Dlink\endcsname~\csname
  2011Sci...331...55Dnote\endcsname

\bibitem[{{De Pontieu} {et~al.}(2007{\natexlab{c}}){De Pontieu}, {McIntosh},
  {Carlsson}, {Hansteen}, {Tarbell}, {Schrijver}, {Title}, {Shine}, {Tsuneta},
  {Katsukawa}, {Ichimoto}, {Suematsu}, {Shimizu}, \&
  {Nagata}}]{2007Sci...318.1574D}
{De Pontieu}, B., {McIntosh}, S.~W., {Carlsson}, M., {et~al.}
  2007{\natexlab{c}}, Science, 318, 1574 \csname
  2007Sci...318.1574Dlink\endcsname~\csname 2007Sci...318.1574Dnote\endcsname

\bibitem[{{De Pontieu} {et~al.}(2014){De Pontieu}, {Title}, {Lemen}, {Kushner},
  {Akin}, {Allard}, {Berger}, {Boerner}, {Cheung}, {Chou}, {Drake}, {Duncan},
  {Freeland}, {Heyman}, {Hoffman}, {Hurlburt}, {Lindgren}, {Mathur}, {Rehse},
  {Sabolish}, {Seguin}, {Schrijver}, {Tarbell}, {W{\"u}lser}, {Wolfson},
  {Yanari}, {Mudge}, {Nguyen-Phuc}, {Timmons}, {van Bezooijen}, {Weingrod},
  {Brookner}, {Butcher}, {Dougherty}, {Eder}, {Knagenhjelm}, {Larsen},
  {Mansir}, {Phan}, {Boyle}, {Cheimets}, {DeLuca}, {Golub}, {Gates}, {Hertz},
  {McKillop}, {Park}, {Perry}, {Podgorski}, {Reeves}, {Saar}, {Testa}, {Tian},
  {Weber}, {Dunn}, {Eccles}, {Jaeggli}, {Kankelborg}, {Mashburn}, {Pust},
  {Springer}, {Carvalho}, {Kleint}, {Marmie}, {Mazmanian}, {Pereira}, {Sawyer},
  {Strong}, {Worden}, {Carlsson}, {Hansteen}, {Leenaarts}, {Wiesmann},
  {Aloise}, {Chu}, {Bush}, {Scherrer}, {Brekke}, {Mart{\'{\i}}nez-Sykora},
  {Lites}, {McIntosh}, {Uitenbroek}, {Okamoto}, {Gummin}, {Auker}, {Jerram},
  {Pool}, \& {Waltham}}]{2014SoPh..289.2733D}
{De Pontieu}, B., {Title}, A.~M., {Lemen}, J.~R., {et~al.} 2014, \solphys, 289,
  2733 \csname 2014SoPh..289.2733Dlink\endcsname~\csname
  2014SoPh..289.2733Dnote\endcsname

\bibitem[{{Foukal}(1971)}]{1971SoPh...20..298F}
{Foukal}, P. 1971, \solphys, 20, 298 \csname
  1971SoPh...20..298Flink\endcsname~\csname 1971SoPh...20..298Fnote\endcsname

\bibitem[{{Golding} {et~al.}(2014){Golding}, {Carlsson}, \&
  {Leenaarts}}]{2014ApJ...784...30G}
{Golding}, T.~P., {Carlsson}, M., \& {Leenaarts}, J. 2014, \apj, 784, 30
  \csname 2014ApJ...784...30Glink\endcsname~\csname
  2014ApJ...784...30Gnote\endcsname

\bibitem[{{Hansteen} {et~al.}(2006){Hansteen}, {De Pontieu}, {Rouppe van der
  Voort}, {van Noort}, \& {Carlsson}}]{2006ApJ...647L..73H}
{Hansteen}, V.~H., {De Pontieu}, B., {Rouppe van der Voort}, L., {van Noort},
  M., \& {Carlsson}, M. 2006, \apjl, 647, L73 \csname
  2006ApJ...647L..73Hlink\endcsname~\csname 2006ApJ...647L..73Hnote\endcsname

\bibitem[{{Henriques}(2012)}]{2012A&A...548A.114H}
{Henriques}, V.~M.~J. 2012, \aap, 548, A114 \csname
  2012A&A...548A.114Hlink\endcsname~\csname 2012A&A...548A.114Hnote\endcsname

\bibitem[{{Judge} {et~al.}(2011){Judge}, {Tritschler}, \& {Chye
  Low}}]{2011ApJ...730L...4J}
{Judge}, P.~G., {Tritschler}, A., \& {Chye Low}, B. 2011, \apjl, 730, L4
  \csname 2011ApJ...730L...4Jlink\endcsname~\csname
  2011ApJ...730L...4Jnote\endcsname

\bibitem[{{Langangen} {et~al.}(2008){Langangen}, {De Pontieu}, {Carlsson},
  {Hansteen}, {Cauzzi}, \& {Reardon}}]{2008ApJ...679L.167L}
{Langangen}, {\O}., {De Pontieu}, B., {Carlsson}, M., {et~al.} 2008, \apjl,
  679, L167 \csname 2008ApJ...679L.167Llink\endcsname~\csname
  2008ApJ...679L.167Lnote\endcsname

\bibitem[{{Leenaarts} {et~al.}(2007){Leenaarts}, {Carlsson}, {Hansteen}, \&
  {Rutten}}]{2007A&A...473..625L}
{Leenaarts}, J., {Carlsson}, M., {Hansteen}, V., \& {Rutten}, R.~J. 2007, \aap,
  473, 625 \csname 2007A&A...473..625Llink\endcsname~\csname
  2007A&A...473..625Lnote\endcsname

\bibitem[{{Leenaarts} {et~al.}(2012){Leenaarts}, {Carlsson}, \& {Rouppe van der
  Voort}}]{2012ApJ...749..136L}
{Leenaarts}, J., {Carlsson}, M., \& {Rouppe van der Voort}, L. 2012, \apj, 749,
  136 \csname 2012ApJ...749..136Llink\endcsname~\csname
  2012ApJ...749..136Lnote\endcsname

\bibitem[{{Leenaarts} {et~al.}(2015){Leenaarts}, {Carlsson}, \& {Rouppe van der
  Voort}}]{2015ApJ...802..136L}
{Leenaarts}, J., {Carlsson}, M., \& {Rouppe van der Voort}, L. 2015, \apj, 802,
  136 \csname 2015ApJ...802..136Llink\endcsname~\csname
  2015ApJ...802..136Lnote\endcsname

\bibitem[{{Lemen} {et~al.}(2012){Lemen}, {Title}, {Akin}, {Boerner}, {Chou},
  {Drake}, {Duncan}, {Edwards}, {Friedlaender}, {Heyman}, {Hurlburt}, {Katz},
  {Kushner}, {Levay}, {Lindgren}, {Mathur}, {McFeaters}, {Mitchell}, {Rehse},
  {Schrijver}, {Springer}, {Stern}, {Tarbell}, {Wuelser}, {Wolfson}, {Yanari},
  {Bookbinder}, {Cheimets}, {Caldwell}, {Deluca}, {Gates}, {Golub}, {Park},
  {Podgorski}, {Bush}, {Scherrer}, {Gummin}, {Smith}, {Auker}, {Jerram},
  {Pool}, {Soufli}, {Windt}, {Beardsley}, {Clapp}, {Lang}, \&
  {Waltham}}]{2012SoPh..275...17L}
{Lemen}, J.~R., {Title}, A.~M., {Akin}, D.~J., {et~al.} 2012, \solphys, 275, 17
  \csname 2012SoPh..275...17Llink\endcsname~\csname
  2012SoPh..275...17Lnote\endcsname

\bibitem[{{Lipartito} {et~al.}(2014){Lipartito}, {Judge}, {Reardon}, \&
  {Cauzzi}}]{2014ApJ...785..109L}
{Lipartito}, I., {Judge}, P.~G., {Reardon}, K., \& {Cauzzi}, G. 2014, \apj,
  785, 109 \csname 2014ApJ...785..109Llink\endcsname~\csname
  2014ApJ...785..109Lnote\endcsname

\bibitem[{{Lockyer}(1868)}]{1868RSPS...17..131L}
{Lockyer}, J.~N. 1868, Proceedings of the Royal Society of London Series I, 17,
  131 \csname 1868RSPS...17..131Llink\endcsname~\csname
  1868RSPS...17..131Lnote\endcsname

\bibitem[{{Mart{\'{\i}}nez-Sykora} {et~al.}(2015){Mart{\'{\i}}nez-Sykora},
  {Rouppe van der Voort}, {Carlsson}, {De Pontieu}, {Pereira}, {Boerner},
  {Hurlburt}, {Kleint}, {Lemen}, {Tarbell}, {Title}, {Wuelser}, {Hansteen},
  {Golub}, {McKillop}, {Reeves}, {Saar}, {Testa}, {Tian}, {Jaeggli}, \&
  {Kankelborg}}]{2015ApJ...803...44M}
{Mart{\'{\i}}nez-Sykora}, J., {Rouppe van der Voort}, L., {Carlsson}, M.,
  {et~al.} 2015, \apj, 803, 44 \csname
  2015ApJ...803...44Mlink\endcsname~\csname 2015ApJ...803...44Mnote\endcsname

\bibitem[{{McIntosh} {et~al.}(2008){McIntosh}, {De Pontieu}, \&
  {Tarbell}}]{2008ApJ...673L.219M}
{McIntosh}, S.~W., {De Pontieu}, B., \& {Tarbell}, T.~D. 2008, \apjl, 673, L219
  \csname 2008ApJ...673L.219Mlink\endcsname~\csname
  2008ApJ...673L.219Mnote\endcsname

\bibitem[{{Meyer} {et~al.}(2012){Meyer}, {Mackay}, \& {van
  Ballegooijen}}]{2012SoPh..278..149M}
{Meyer}, K.~A., {Mackay}, D.~H., \& {van Ballegooijen}, A.~A. 2012, \solphys,
  278, 149 \csname 2012SoPh..278..149Mlink\endcsname~\csname
  2012SoPh..278..149Mnote\endcsname

\bibitem[{{Noyes}(1967)}]{1967IAUS...28..293N}
{Noyes}, R.~W. 1967, in IAU Symposium, Vol.~28, Aerodynamic Phenomena in
  Stellar Atmospheres, ed. R.~N. {Thomas}, 293 \csname
  1967IAUS...28..293Nlink\endcsname~\csname 1967IAUS...28..293Nnote\endcsname

\bibitem[{{Pereira} {et~al.}(2012){Pereira}, {De Pontieu}, \&
  {Carlsson}}]{2012ApJ...759...18P}
{Pereira}, T.~M.~D., {De Pontieu}, B., \& {Carlsson}, M. 2012, \apj, 759, 18
  \csname 2012ApJ...759...18Plink\endcsname~\csname
  2012ApJ...759...18Pnote\endcsname

\bibitem[{{Pereira} {et~al.}(2014){Pereira}, {De Pontieu}, {Carlsson},
  {Hansteen}, {Tarbell}, {Lemen}, {Title}, {Boerner}, {Hurlburt}, {W{\"u}lser},
  {Mart{\'{\i}}nez-Sykora}, {Kleint}, {Golub}, {McKillop}, {Reeves}, {Saar},
  {Testa}, {Tian}, {Jaeggli}, \& {Kankelborg}}]{2014ApJ...792L..15P}
{Pereira}, T.~M.~D., {De Pontieu}, B., {Carlsson}, M., {et~al.} 2014, \apjl,
  792, L15 \csname 2014ApJ...792L..15Plink\endcsname~\csname
  2014ApJ...792L..15Pnote\endcsname

\bibitem[{{Pesnell} {et~al.}(2012){Pesnell}, {Thompson}, \&
  {Chamberlin}}]{2012SoPh..275....3P}
{Pesnell}, W.~D., {Thompson}, B.~J., \& {Chamberlin}, P.~C. 2012, \solphys,
  275, 3 \csname 2012SoPh..275....3Plink\endcsname~\csname
  2012SoPh..275....3Pnote\endcsname

\bibitem[{{Robustini} {et~al.}(2016){Robustini}, {Leenaarts}, {de la Cruz
  Rodrigu{\'{e}}z}, \& {Rouppe van der Voort}}]{2016A&A...590A..57R}
{Robustini}, C., {Leenaarts}, J., {de la Cruz Rodrigu{\'{e}}z}, J., \& {Rouppe
  van der Voort}, L. 2016, \aap, 590, A57 \csname
  2016A&A...590A..57Rlink\endcsname~\csname 2016A&A...590A..57Rnote\endcsname

\bibitem[{{Rouppe van der Voort} \& {de la Cruz
  Rodr{\'{\i}}guez}(2013)}]{2013ApJ...776...56R}
{Rouppe van der Voort}, L. \& {de la Cruz Rodr{\'{\i}}guez}, J. 2013, \apj,
  776, 56 \csname 2013ApJ...776...56Rlink\endcsname~\csname
  2013ApJ...776...56Rnote\endcsname

\bibitem[{{Rouppe van der Voort} {et~al.}(2015){Rouppe van der Voort}, {De
  Pontieu}, {Pereira}, {Carlsson}, \& {Hansteen}}]{2015ApJ...799L...3R}
{Rouppe van der Voort}, L., {De Pontieu}, B., {Pereira}, T.~M.~D., {Carlsson},
  M., \& {Hansteen}, V. 2015, \apjl, 799, L3 \csname
  2015ApJ...799L...3Rlink\endcsname~\csname 2015ApJ...799L...3Rnote\endcsname

\bibitem[{{Rouppe van der Voort} {et~al.}(2009){Rouppe van der Voort},
  {Leenaarts}, {De Pontieu}, {Carlsson}, \& {Vissers}}]{2009ApJ...705..272R}
{Rouppe van der Voort}, L., {Leenaarts}, J., {De Pontieu}, B., {Carlsson}, M.,
  \& {Vissers}, G. 2009, \apj, 705, 272 \csname
  2009ApJ...705..272Rlink\endcsname~\csname 2009ApJ...705..272Rnote\endcsname

\bibitem[{{Rouppe van der Voort} {et~al.}(2007){Rouppe van der Voort}, {De
  Pontieu}, {Hansteen}, {Carlsson}, \& {van Noort}}]{2007ApJ...660L.169R}
{Rouppe van der Voort}, L.~H.~M., {De Pontieu}, B., {Hansteen}, V.~H.,
  {Carlsson}, M., \& {van Noort}, M. 2007, \apjl, 660, L169 \csname
  2007ApJ...660L.169Rlink\endcsname~\csname 2007ApJ...660L.169Rnote\endcsname

\bibitem[{{Rutten}(2016)}]{2016A&A...590A.124R}
{Rutten}, R.~J. 2016, \aap, 590, A124 \csname
  2016A&A...590A.124Rlink\endcsname~\csname 2016A&A...590A.124Rnote\endcsname

\bibitem[{{Rutten} {et~al.}(2011){Rutten}, {Leenaarts}, {Rouppe van der Voort},
  {De Wijn}, {Carlsson}, \& {Hansteen}}]{2011A&A...531A..17R}
{Rutten}, R.~J., {Leenaarts}, J., {Rouppe van der Voort}, L.~H.~M., {et~al.}
  2011, \aap, 531, A17 \csname 2011A&A...531A..17Rlink\endcsname~\csname
  2011A&A...531A..17Rnote\endcsname

\bibitem[{{Rutten} {et~al.}(2008){Rutten}, {van Veelen}, \&
  {S{\"u}tterlin}}]{2008SoPh..251..533R}
{Rutten}, R.~J., {van Veelen}, B., \& {S{\"u}tterlin}, P. 2008, \solphys, 251,
  533 \csname 2008SoPh..251..533Rlink\endcsname~\csname
  2008SoPh..251..533Rnote\endcsname

\bibitem[{{S{\'a}nchez-Andrade Nu{\~n}o} {et~al.}(2008){S{\'a}nchez-Andrade
  Nu{\~n}o}, {Bello Gonz{\'a}lez}, {Blanco Rodr{\'{\i}}guez}, {Kneer}, \&
  {Puschmann}}]{2008A&A...486..577S}
{S{\'a}nchez-Andrade Nu{\~n}o}, B., {Bello Gonz{\'a}lez}, N., {Blanco
  Rodr{\'{\i}}guez}, J., {Kneer}, F., \& {Puschmann}, K.~G. 2008, \aap, 486,
  577 \csname 2008A&A...486..577Slink\endcsname~\csname
  2008A&A...486..577Snote\endcsname

\bibitem[{{Scharmer} {et~al.}(2003){Scharmer}, {Bjelksjo}, {Korhonen},
  {Lindberg}, \& {Petterson}}]{2003SPIE.4853..341S}
{Scharmer}, G.~B., {Bjelksjo}, K., {Korhonen}, T.~K., {Lindberg}, B., \&
  {Petterson}, B. 2003, in \procspie, Vol. 4853, Innovative Telescopes and
  Instrumentation for Solar Astrophysics, ed. S.~L. {Keil} \& S.~V. {Avakyan},
  341--350 \csname 2003SPIE.4853..341Slink\endcsname~\csname
  2003SPIE.4853..341Snote\endcsname

\bibitem[{{Scharmer} {et~al.}(2008){Scharmer}, {Narayan}, {Hillberg}, {de la
  Cruz Rodrigu{\'{e}}z}, {L{\"o}fdahl}, {Kiselman}, {S{\"u}tterlin}, {van
  Noort}, \& {Lagg}}]{2008ApJ...689L..69S}
{Scharmer}, G.~B., {Narayan}, G., {Hillberg}, T., {et~al.} 2008, \apjl, 689,
  L69 \csname 2008ApJ...689L..69Slink\endcsname~\csname
  2008ApJ...689L..69Snote\endcsname

\bibitem[{{Sekse} {et~al.}(2012){Sekse}, {Rouppe van der Voort}, \& {De
  Pontieu}}]{2012ApJ...752..108S}
{Sekse}, D.~H., {Rouppe van der Voort}, L., \& {De Pontieu}, B. 2012, \apj,
  752, 108 \csname 2012ApJ...752..108Slink\endcsname~\csname
  2012ApJ...752..108Snote\endcsname

\bibitem[{{Sekse} {et~al.}(2013){Sekse}, {Rouppe van der Voort}, {De Pontieu},
  \& {Scullion}}]{2013ApJ...769...44S}
{Sekse}, D.~H., {Rouppe van der Voort}, L., {De Pontieu}, B., \& {Scullion}, E.
  2013, \apj, 769, 44 \csname 2013ApJ...769...44Slink\endcsname~\csname
  2013ApJ...769...44Snote\endcsname

\bibitem[{{Shine} {et~al.}(1994){Shine}, {Title}, {Tarbell}, {Smith}, {Frank},
  \& {Scharmer}}]{1994ApJ...430..413S}
{Shine}, R.~A., {Title}, A.~M., {Tarbell}, T.~D., {et~al.} 1994, \apj, 430, 413
  \csname 1994ApJ...430..413Slink\endcsname~\csname
  1994ApJ...430..413Snote\endcsname

\bibitem[{{Skogsrud} {et~al.}(2015){Skogsrud}, {Rouppe van der Voort}, {De
  Pontieu}, \& {Pereira}}]{2015ApJ...806..170S}
{Skogsrud}, H., {Rouppe van der Voort}, L., {De Pontieu}, B., \& {Pereira},
  T.~M.~D. 2015, \apj, 806, 170 \csname
  2015ApJ...806..170Slink\endcsname~\csname 2015ApJ...806..170Snote\endcsname

\bibitem[{{van Noort} {et~al.}(2005){van Noort}, {Rouppe van der Voort}, \&
  {L{\"o}fdahl}}]{2005SoPh..228..191V}
{van Noort}, M., {Rouppe van der Voort}, L., \& {L{\"o}fdahl}, M.~G. 2005,
  \solphys, 228, 191 \csname 2005SoPh..228..191Vlink\endcsname~\csname
  2005SoPh..228..191Vnote\endcsname

\bibitem[{{Vissers} \& {Rouppe van der Voort}(2012)}]{2012ApJ...750...22V}
{Vissers}, G. \& {Rouppe van der Voort}, L. 2012, \apj, 750, 22 \csname
  2012ApJ...750...22Vlink\endcsname~\csname 2012ApJ...750...22Vnote\endcsname

\bibitem[{{Vissers} {et~al.}(2015){Vissers}, {Rouppe van der Voort}, {Rutten},
  {Carlsson}, \& {De Pontieu}}]{2015ApJ...812...11V}
{Vissers}, G.~J.~M., {Rouppe van der Voort}, L.~H.~M., {Rutten}, R.~J.,
  {Carlsson}, M., \& {De Pontieu}, B. 2015, \apj, 812, 11 \csname
  2015ApJ...812...11Vlink\endcsname~\csname 2015ApJ...812...11Vnote\endcsname

\bibitem[{{Wilhelm} {et~al.}(1995){Wilhelm}, {Curdt}, {Marsch}, {Sch{\"u}hle},
  {Lemaire}, {Gabriel}, {Vial}, {Grewing}, {Huber}, {Jordan}, {Poland},
  {Thomas}, {K{\"u}hne}, {Timothy}, {Hassler}, \&
  {Siegmund}}]{1995SoPh..162..189W}
{Wilhelm}, K., {Curdt}, W., {Marsch}, E., {et~al.} 1995, \solphys, 162, 189
  \csname 1995SoPh..162..189Wlink\endcsname~\csname
  1995SoPh..162..189Wnote\endcsname

\bibitem[{{Yurchyshyn} {et~al.}(2014){Yurchyshyn}, {Abramenko}, {Kosovichev},
  \& {Goode}}]{2014ApJ...787...58Y}
{Yurchyshyn}, V., {Abramenko}, V., {Kosovichev}, A., \& {Goode}, P. 2014, \apj,
  787, 58 \csname 2014ApJ...787...58Ylink\endcsname~\csname
  2014ApJ...787...58Ynote\endcsname

\end{thebibliography}

\end{document}